\documentclass[9pt,shortpaper,twoside,web]{ieeecolor}
\usepackage{generic}
\usepackage{cite}
\usepackage{amsmath,amssymb,amsfonts}
\usepackage{algorithmic}
\usepackage{graphicx}
\usepackage{textcomp}
\usepackage{longtable}
\usepackage{cuted}
\usepackage{booktabs}
\usepackage{amssymb}
\usepackage{array}       % 增强表格功能
\usepackage{multirow}    % 支持单元格合并
\usepackage{color}
\usepackage[table]{xcolor} % 启用表格着色支持
\usepackage{colortbl}      % 提供表格相关颜色功能
\usepackage{tabularx}
\usepackage{url}
\usepackage{arydshln}

\usepackage{tikz}
\usetikzlibrary{arrows.meta, shadows}  % 箭头与阴影
\usepackage{forest}
\useforestlibrary{edges}        % forest增强功能
\usepackage{rotating}                  % 支持 rotatebox

\definecolor{mypurple}{RGB}{166, 146, 195}
\definecolor{darkpurple}{RGB}{102, 8, 116}
\usepackage{hyperref}
\hypersetup{
    colorlinks=true,
    linkcolor=darkpurple,
    urlcolor=darkpurple,
    citecolor=darkpurple,
}

\def\BibTeX{{\rm B\kern-.05em{\sc i\kern-.025em b}\kern-.08em
    T\kern-.1667em\lower.7ex\hbox{E}\kern-.125emX}}
\begin{document}
\bstctlcite{IEEEexample:BSTcontrol} % control ref author number

% \title{Generative Models in Computational Pathology: A Comprehensive Survey on Methods, Applications, and Challenges}

\title{Content Generation Models in Computational Pathology: A Comprehensive Survey on Methods, Applications, and Challenges}

\author{Yuan Zhang,
        Xinfeng Zhang, 
        Xiaoming Qi,
        Xinyu Wu,
        Feng Chen,\\ 
        Guanyu Yang, \IEEEmembership{Senior Member, IEEE},
        and Huazhu Fu, \IEEEmembership{Senior Member, IEEE}
\thanks{Yuan Zhang and Xinfeng Zhang contributed equally to this work. Corresponding author: Guanyu Yang and Huazhu Fu.}    
\thanks{Y.~Zhang and G.~Yang are with the Key Laboratory of New Generation Artificial Intelligence Technology and Its Interdisciplinary Applications (Southeast University), Ministry of Education, Nanjing, China (e-mail:
yuanzhang\_@seu.edu.cn, yang.list@seu.edu.cn).}  
\thanks{X.~Zhang is with the School of Biomedical Engineering, Tsinghua University, Beijing, China (e-mail: zhang-xf22@mails.tsinghua.edu.cn).}  
\thanks{X.~Qi is with the Department of Biomedical Engineering and Department of Electrical and Computer Engineering, National University of Singapore, Singapore (e-mail: qixiaoming12138@163.com).}  
\thanks{X.~Wu is with the School of Information Science and Engineering, Southeast University, Nanjing, China (e-mail: xinyuwu@seu.edu.cn).}  
\thanks{F.~Chen is with the Department of Biostatistics, Center for Global Health, School of Public Health, Nanjing Medical University, Nanjing, China (e-mail: fengchen@njmu.edu.cn).} 
\thanks{H.~Fu is with the Institute of High-Performance Computing, Agency for Science, Technology and Research, Singapore (e-mail: hzfu@ieee.org).}  
        }

% \author{First A. Author, \IEEEmembership{Fellow, IEEE}, Second B. Author, and Third C. Author, Jr., \IEEEmembership{Member, IEEE}
% \thanks{This paragraph of the first footnote will contain the date on 
% which you submitted your paper for review. It will also contain support 
% information, including sponsor and financial support acknowledgment. For 
% example, ``This work was supported in part by the U.S. Department of 
% Commerce under Grant BS123456.'' }
% \thanks{The next few paragraphs should contain 
% the authors' current affiliations, including current address and e-mail. For 
% example, F. A. Author is with the National Institute of Standards and 
% Technology, Boulder, CO 80305 USA (e-mail: author@boulder.nist.gov). }
% \thanks{S. B. Author, Jr., was with Rice University, Houston, TX 77005 USA. He is 
% now with the Department of Physics, Colorado State University, Fort Collins, 
% CO 80523 USA (e-mail: author@lamar.colostate.edu).}
% \thanks{T. C. Author is with 
% the Electrical Engineering Department, University of Colorado, Boulder, CO 
% 80309 USA, on leave from the National Research Institute for Metals, 
% Tsukuba, Japan (e-mail: author@nrim.go.jp).}
% }

\maketitle

\begin{abstract}
\textcolor{black}{Content generation modeling} has emerged as a promising direction in computational pathology, offering capabilities such as data-efficient learning, synthetic data augmentation, and task-oriented generation across diverse diagnostic tasks. This review provides a comprehensive synthesis of recent progress in the field, organized into four key domains: image generation, text generation, molecular profile–morphology generation, and other specialized generation applications. By analyzing over 150 representative studies, we trace the evolution of \textcolor{black}{content generation architectures}---from early generative adversarial networks to recent advances in diffusion models and generative vision–language models. We further examine the datasets and evaluation protocols commonly used in this domain and highlight ongoing limitations, including challenges in generating high-fidelity whole slide images, clinical interpretability, and concerns related to the ethical and legal implications of synthetic data. The review concludes with a discussion of open challenges and prospective research directions, with an emphasis on developing integrated and clinically deployable generation systems. This work aims to provide a foundational reference for researchers and practitioners developing \textcolor{black}{content generation models} in computational pathology.

\end{abstract}


\section{Introduction}
\label{sec:introduction}
\begin{figure}[!t]
\centerline{\includegraphics[width=0.4\textwidth]{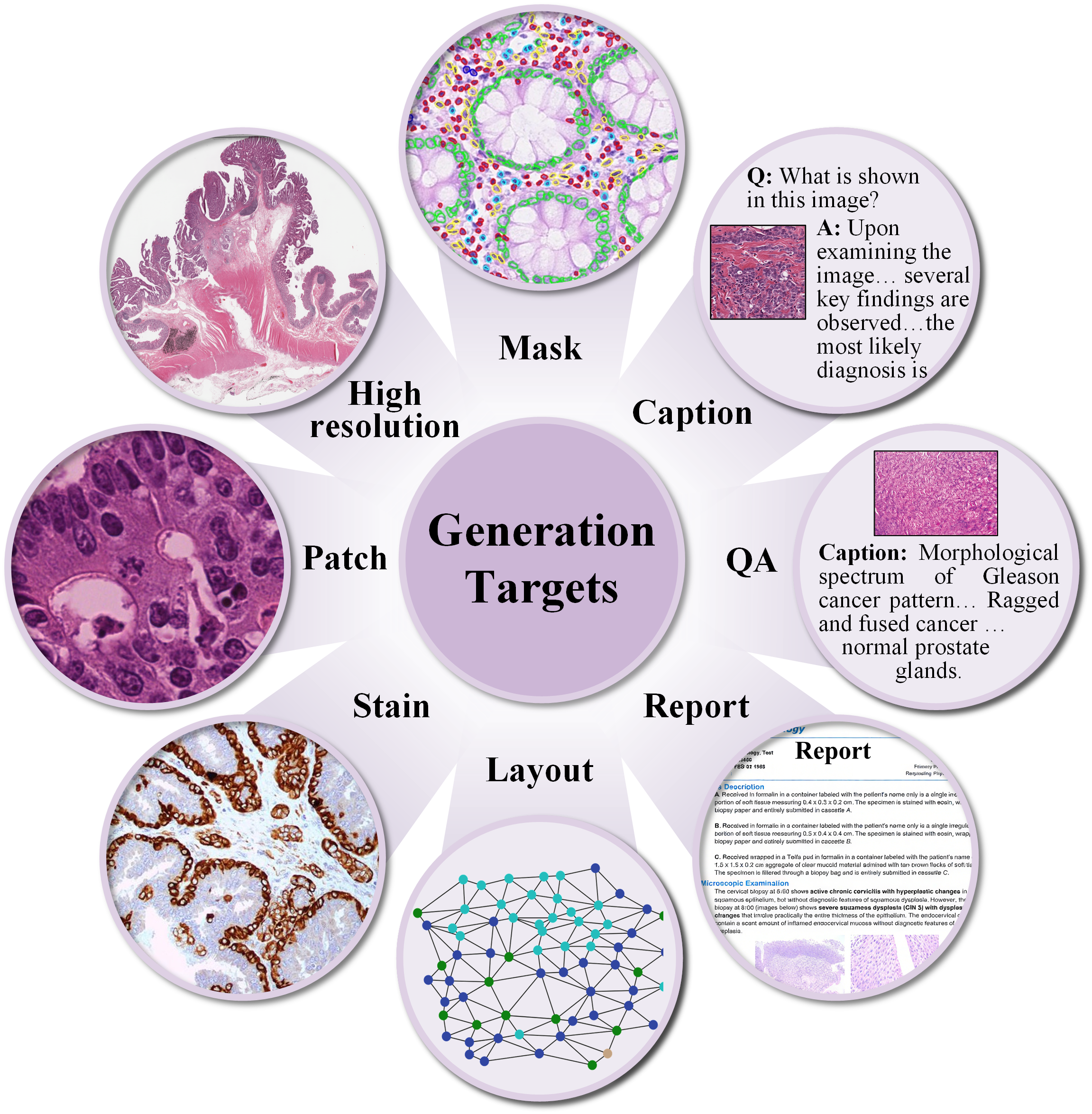}}
\caption{Diverse generation targets in computational pathology. This figure illustrates the spectrum of generative tasks, covering image and mask synthesis, spatial layout generation, textual report generation, and visual question answering.}
\label{fig_sun}
\end{figure}

\textcolor{black}{Content generation models are machine learning models that learn underlying data distributions and generate new biomedical instances that preserve the essential characteristics of the training data.} They have become a transformative force in medical image analysis, offering capabilities that extend beyond traditional discriminative learning \cite{van2024syntheticdata,zhang2022shifting}. By modeling complex data distributions to synthesize high-fidelity content, these models enable a wide range of downstream tasks, such as data augmentation, cross-modal translation, and representation learning \cite{kazerouni2023diffusionsurvey}. \textcolor{black}{The diversity of these generation targets in computational pathology is illustrated in Fig.~\ref{fig_sun}.} Their utility is especially pronounced in computational pathology \cite{deshpande2023thesis}, a domain characterized by high-resolution whole-slide images (WSIs), heterogeneous tissue morphology, and significant annotation burdens.

\begin{figure}[t]
\centerline{\includegraphics[width=0.45\textwidth]{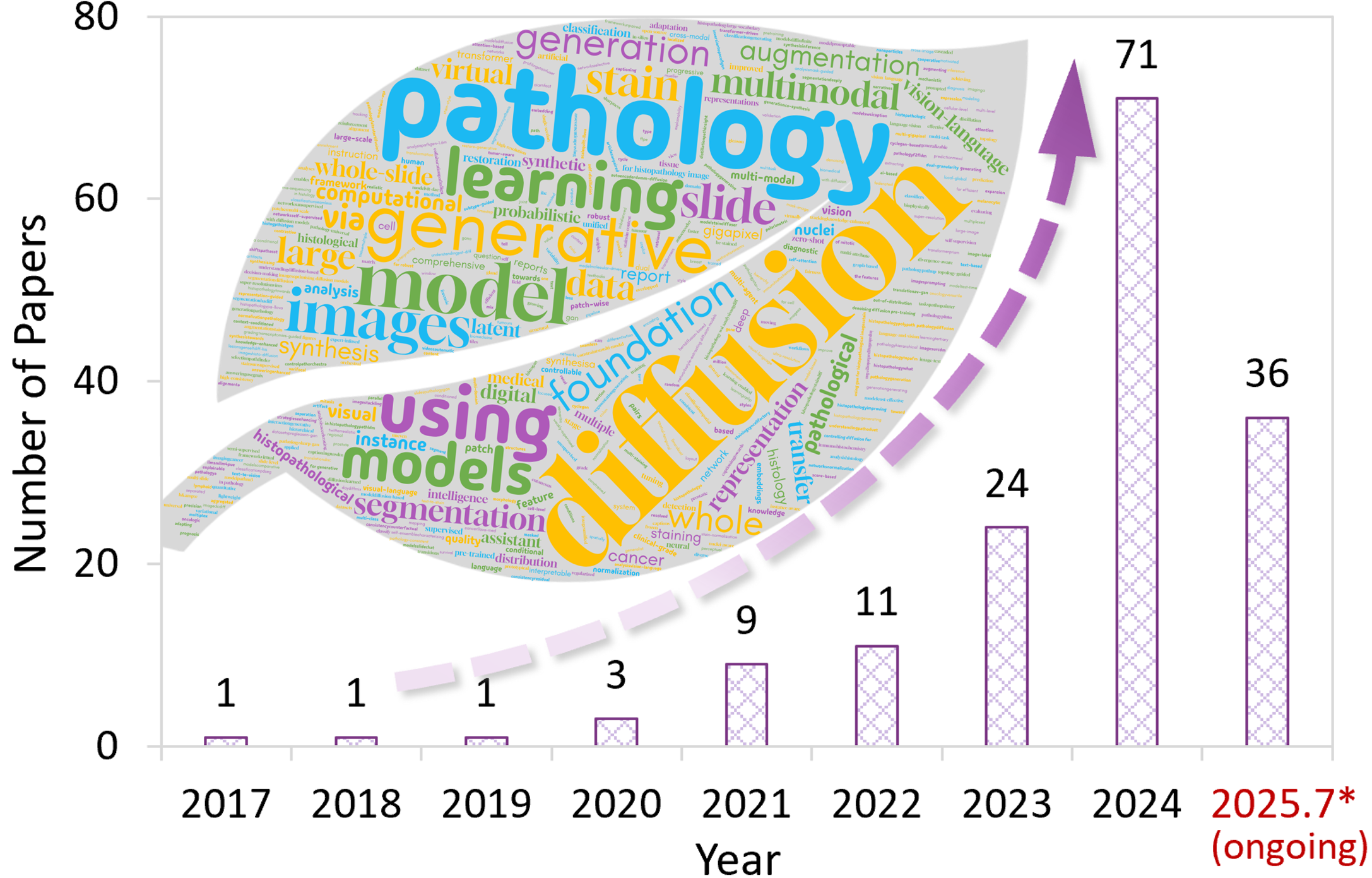}}
\caption{\color{black}Number of published papers on generative models in pathology from 2017 to July 2025. The red label “2025.7* (ongoing)” marks ongoing work up to July, while the purple curve shows the exponential growth trend. The accompanying word cloud highlights the most frequent terms, extracted from the titles of the collected articles.}
\label{num}
\end{figure}

Computational pathology has rapidly evolved into a data-intensive field driven by advances in deep learning and whole slide imaging \cite{song2023artificial}. However, this progress also reveals core limitations in data availability, generalization, and robustness across clinical domains \cite{croitoru2023diffusionsurvey}. Generative models such as generative adversarial networks (GAN) \textcolor{black}{\cite{fu2024pix2path, howard2024generative, 10160043, Wang_2025_CVPR}}, variational autoencoders (VAE) \textcolor{black}{\cite{wang2024cross-modal, wood2025genst, wan2023integrating, boyd2021self}}, diffusion models \textcolor{black}{\cite{NEURIPS2023_f64927f5, he2023artifact, 10.1007/978-3-031-43987-2_53}}, and foundation models like large language models (LLM) and vision-language models (VLM) \textcolor{black}{\cite{Lu2024, Chen_2025_CVPR, Seyfioglu2024quilt-llava}} have proven effective in addressing these challenges. These models support a wide range of applications, including synthetic data generation, stain normalization, style transfer, tissue synthesis, and rare pathology simulation, thereby enhancing learning-based approaches under data-scarce or domain-shifted conditions. Research on \textcolor{black}{content generation models} in pathology has grown rapidly, as illustrated in Fig.~\ref{num}. After a period of sparse, exploratory studies, publication volume increased steadily, with a sharp rise from 2024 onward. This inflection point marks a broader shift in the field, as researchers increasingly recognize the unique advantages that \textcolor{black}{content generation models} offer in addressing long-standing challenges in digital pathology, including limited data availability, high annotation costs, and the demand for high-fidelity, consistent synthetic images.

Despite the rapid proliferation of \textcolor{black}{content generation approaches} in pathology, a focused review that systematically organizes methods and assesses their impact is still lacking. Prior surveys in computer vision often examine \textcolor{black}{content generation models} from broad or algorithmic perspectives \cite{he2025diffusionpami,guo2024diffusionbiology}, leaving pathology-specific challenges underexplored. \textcolor{black}{Notably, Chanda et al. (2024) \cite{chanda2024new} present a broad catalog of foundation and vision-language models,} \textcolor{black}{where generative tasks are listed as one among many applications, whereas Bilal et al. (2025) \cite{bilal2025foundation} concentrate on evaluation frameworks for foundation models rather than their pathology-specific use. In contrast, our review is centered on content generation models as a distinct methodological paradigm.} Given the unique demands of pathology, such as cell-level structural fidelity, high-resolution WSIs, and stain consistency, there is a pressing need to consolidate current developments under a pathology-centric lens. With the advent of foundation models and cross-modal generation strategies, the \textcolor{black}{content generation paradigm} in pathology is undergoing a significant shift, warranting timely reflection and synthesis.

% tree
\begin{figure}[htbp]
    \centering
    \begin{flushright}
\begin{minipage}{0.5\textwidth}
\centering
\begin{forest}
for tree={
    grow=east, reversed,
    % 注释掉draw，移除方框的线
    % draw,
    rounded corners=3pt,
    anchor=west,
    parent anchor=east,
    child anchor=west,
    align=left,
    l sep=15pt,
    s sep=6pt,
    edge={-, thick, mypurple},
    edge path={
        \noexpand\path[\forestoption{edge}, rounded corners=2pt]
        (!u.parent anchor) -- ([xshift=10pt]!u.parent anchor) |- (.child anchor);
    },
    font=\scriptsize,
    where level=1{fill=mypurple!40}{}, % 为 Level 1 节点上色
    where level=2{fill=mypurple!15}{}, % 为 Level 2 节点上色
    for descendants={text width=0.45\textwidth},
    }
% 根节点
[{\rotatebox{90}{\centering \textbf{Content Generation Model in Pathology}}},fill=mypurple, text=black
  [{\textbf{Image Generation} (\ref{sec:image_generation})}, tier=level1, text width=0.32\textwidth
    % 第二层子节点
    [{Synthetic Image/Augmentation (\ref{sec:synthetic_image})}, tier=level2]
    [{Mask-Guided Generation (\ref{sec:mask-guided})}, tier=level2]
    [{Artifact Restoration (\ref{sec:artifacts_restoration})}, tier=level2]
    [{High/Multiple-Resolution (\ref{sec:high/multiple-resolution})}, tier=level2]
    [{Text to Image(\ref{sec:text_to_image})}, tier=level2]
    [{Stain Synthesis(\ref{sec:stain_synthesis})}, tier=level2]    
  ]
  [{\textbf{Text Generation} (\ref{sec:text_generation})}, tier=level1, text width=0.32\textwidth
    [{Image Captioning (\ref{sec:image_captioning})}, tier=level2]
    [{Question Answering (\ref{sec:question_answering})}, tier=level2]
    [{Report Generation (\ref{sec:report_generation})}, tier=level2]
    [{Report Abstraction (\ref{sec:report_abstraction})}, tier=level2]
  ]
  [{\textbf{Molecular Profiles-Morpho-} \\ \textbf{logy Generation} (\ref{sec:molecular_profiles})}, tier=level1, text width=0.32\textwidth
    [{Virtual Molecular Profiling (\ref{sec:virtual_molecular_profiling})}, tier=level2]
    [{Reverse Morphology Generation (\ref{sec:reserve_morphology_generation})}, tier=level2]
  ]
  [{\textbf{Other Generation} (\ref{sec:other_generation})}, tier=level1, text width=0.32\textwidth
    [{Spatial Layout (\ref{sec:spacial_layout})}, tier=level2]
    % [{Molecular/Genomic (\ref{sec:other2})}, tier=level2]
    [{Semantic Output (\ref{sec:semantic_output})}, tier=level2]
    [{Latent Representation(\ref{sec:latent_representation})}, tier=level2]
    [{Cell Simulation (\ref{sec:cell_simulation})}, tier=level2]
  ]
]
\end{forest}
\end{minipage}
\end{flushright}
    \caption{\color{black}{Taxonomy of content generation models in computational pathology.}}
    \label{fig:fourtask}
\end{figure}

\color{black}
In this survey, we present a comprehensive and systematic overview of \textcolor{black}{content generation models} in computational pathology. 
We analyze over 150 peer-reviewed papers encompassing diverse generation targets, \textcolor{black}{with the data sources and search strategy detailed in Supplementary Materials Section II.}
To facilitate a structured understanding of this rapidly evolving field, we propose a multi-dimensional taxonomy that organizes prior work by generation targets, algorithmic approaches, application tasks, and datasets. This taxonomy delineates four major categories: image generation, text generation, molecular profile-morphology generation, and other specialized tasks. Within these categories, representative applications include synthetic image augmentation, mask- or text-guided synthesis, stain style transfer, image captioning and report generation, virtual molecular profiling, and spatial or semantic layout generation, among others.
\textcolor{black}{The scope of this review is limited to models with generative capacity, defined by the presence of a decoder that can synthesize new images or text. In contrast, encoder-only or contrastive foundation models, such as CLIP \cite{radford2021learning}, ALIGN \cite{jia2021scaling}, or DINOv2 \cite{oquab2023dinov2}, are considered non-generative, as they learn aligned representations or similarity scores but do not create new content.}
We further review representative modeling paradigms, including GANs, VAEs, diffusion models, and large language and vision–language models. Their historical development is presented in chronological order, as shown in Fig.~\ref{fig:timeline}. In addition, we compile information on widely used datasets, providing a practical resource for researchers entering the field. Finally, the Discussion synthesizes demonstrated capabilities and impact, analyzes critical limitations and risks, and outlines future directions toward trustworthy clinical adoption.

Our contributions are as follows:
\begin{itemize}
\item We provide the first comprehensive survey of \textcolor{black}{content generation models} in computational pathology, covering more than 150 publications up to July 2025.
\item We propose a unified taxonomy that organizes existing studies by generation targets and modeling approaches, offering a structured view of methodological advances in pathology.
\item We curate commonly used datasets across major generation tasks and critically discuss open challenges and future directions.
\end{itemize}
The remainder of this survey is organized as follows. Section~\ref{sec:models} reviews representative generative models in a historical context. Section~\ref{sec:tasks} categorizes applications across image, text, molecular profile-morphology, and other emerging generation tasks. Section~\ref{sec:datasets} summarizes commonly used datasets. Section~\ref{sec:discussion} discusses capabilities, limitations, and pathways toward clinical deployment.

\section{Generation Model}
\label{sec:models}

\begin{figure*}[!t]  
    \centering   
    \includegraphics[width=0.9\textwidth]{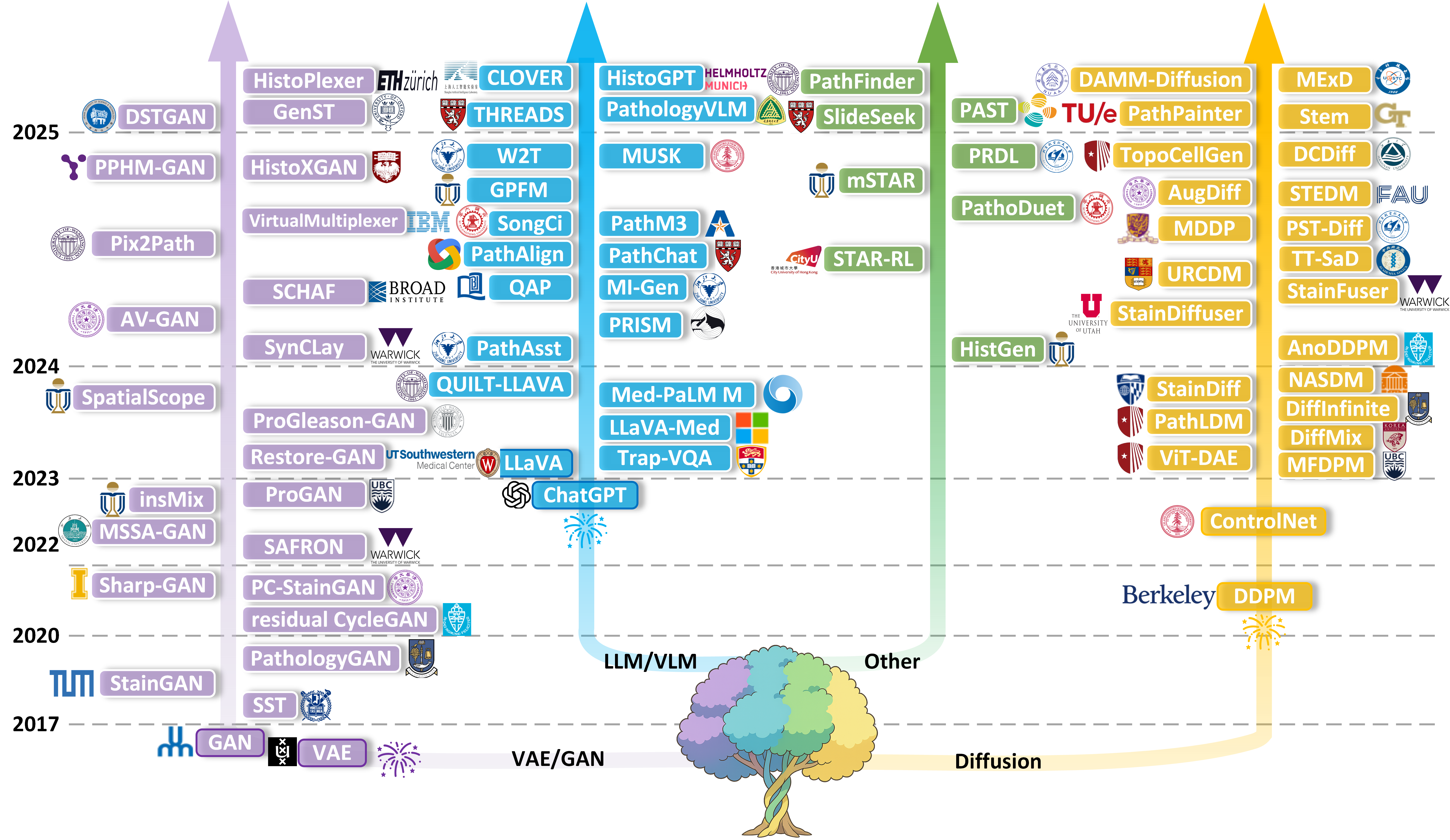}  
    % \caption{\textcolor{black}{A comprehensive visual timeline of the development trajectory of content generation models in pathology from 2017 to 2025.}}
    \caption{Timeline of key developments in pathological generation (2017-2025). This timeline illustrates the major milestones and developmental trajectory of generative models across methods in pathology.}
    \label{fig:timeline}  
\end{figure*} 

\begin{figure*}[!t]  
    \centering   
    \includegraphics[width=\textwidth]{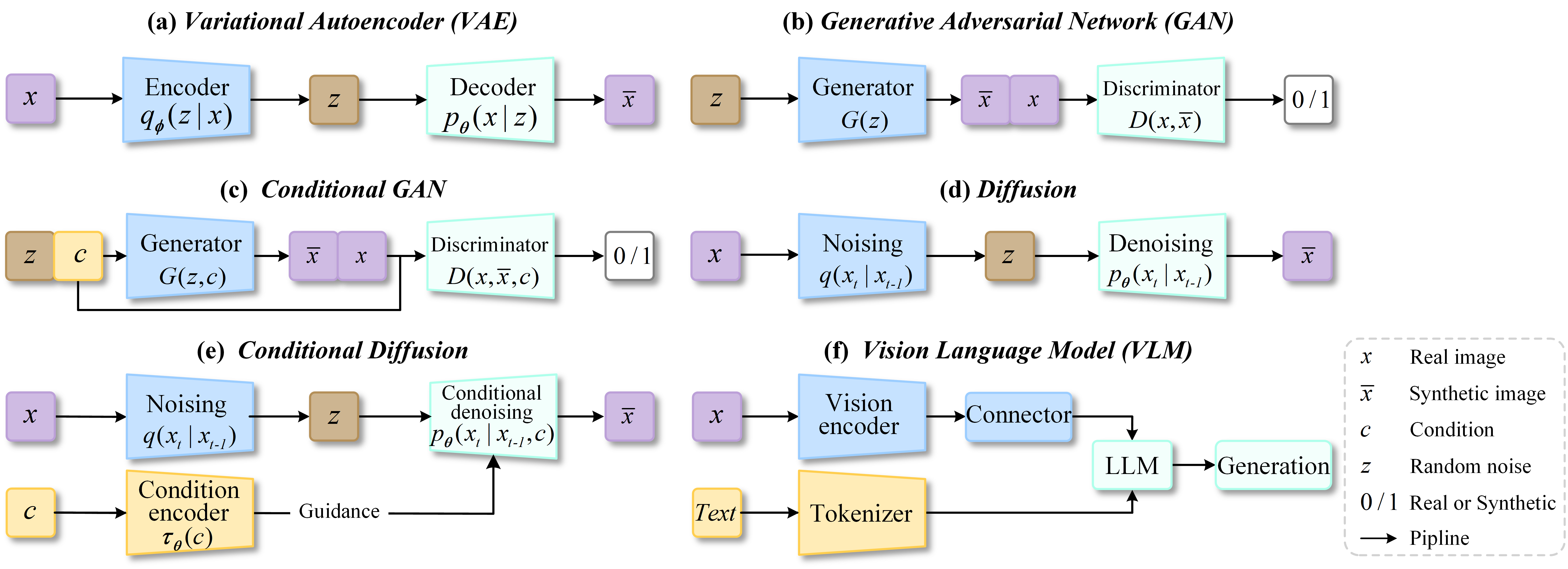}  
    \caption{Schematics of representative generative model architectures. (a) VAE learns a probabilistic latent space and reconstructs images through encoder-decoder optimization. (b) GAN consists of a generator and a discriminator trained in a minimax game to produce realistic outputs. (c) Conditional GAN introduces external conditioning to guide the generation process. (d) The diffusion model performs iterative denoising to generate data by learning the reverse of a noise injection process. (e) Conditional Diffusion extends the diffusion framework with conditioning inputs to steer the generative trajectory during denoising. (f) VLM fuses image and text inputs through an LLM for generation, supporting diverse multimodal tasks.}
    % \caption{Schematics of Key Generative Model Architectures. This figure outlines the core principles of major architectures: (a) VAE for latent space reconstruction; (b) GAN with its generator-discriminator dynamic; (c) Conditional GAN guided by external inputs; (d) Diffusion Models reversing a noising process; (e) Conditional Diffusion steered by guidance; and (f) VLM fusing multimodal inputs.}
    \label{fig:models}  
\end{figure*}

\textcolor{black}{Content generation models in computational pathology have undergone a rapid evolution over the past decade, marked by successive paradigm shifts driven by architectural breakthroughs. This trajectory is summarized in the timeline shown in Figure~\ref{fig:timeline} and further illustrated by representative model families in Figure~\ref{fig:models}. The field's foundations were established around 2014 by \textbf{Variational Autoencoders} and \textbf{Generative Adversarial Networks}, which pioneered the synthesis of images from learned latent spaces. A subsequent shift occurred around 2020 with the advent of \textbf{Diffusion Models}, enabling a new level of high-fidelity image synthesis through an iterative denoising process. More recently, since 2022, the rise of \textbf{Large Language and Vision-Language Models} has reoriented the field from purely pixel-level tasks towards complex, multimodal reasoning and text generation. Beyond these three mainstream paradigms, a set of other distinct approaches, including novel hybrid architectures and advanced training strategies, is also being explored. In the} \textcolor{black}{ following sections, we describe these model families, focusing on their algorithmic innovations and key applications within pathology.}

\subsection{VAE and GAN-based Models}
\textcolor{black}{Variational Autoencoders~\cite{Kingma_2019} introduced the idea of encoding data into a latent distribution and reconstructing it through probabilistic decoding. While VAEs typically generate blurrier images compared} \textcolor{black}{to GANs or diffusion models, their latent structure is well-suited to representation learning~\cite{boyd2021self} and conditional data synthesis~\cite{Zhou2025RobustMS}. In pathology, VAEs have been used for self-supervised pretraining and feature-space augmentation for classification tasks~\cite{10643565}.}

\textcolor{black}{Generative Adversarial Networks~\cite{NIPS2014_goodfellow}, proposed in 2014, introduced adversarial training between a generator and a discriminator to produce realistic samples. Conditional variants such as cGAN~\cite{mirza2014conditional}, Pix2Pix~\cite{isola2017pix2pix}, and CycleGAN~\cite{bouteldja2022tackling} advanced GANs toward controlled, paired, and unpaired image-to-image translation. These architectures have been instrumental in stain normalization~\cite{zanjani2018stain, breen2024generative}, virtual staining~\cite{Kawai2024Vmia, liu2021unpaired}, and cross-domain adaptation~\cite{shen2023federated} in histopathology. Later models such as StyleGAN~\cite{karras2019stylegan} added style-based modulation to the latent space, enabling fine-grained control of image attributes. In computational pathology, StyleGAN and its variants have supported mitotic figure synthesis~\cite{bahadir2023characterizing}, rare-cell augmentation~\cite{golfe2023progleason}, and attribute-conditioned generation~\cite{Cosynthesis2024eccv}. The discriminator components in GANs have also been repurposed for unsupervised learning of biologically meaningful features~\cite{8402089}, making GANs a versatile tool for both generation and representation learning.}

\subsection{Diffusion-based Models}
\textcolor{black}{Diffusion models generate images by learning to reverse a stochastic noising process. This class of models, beginning with the Denoising Diffusion Probabilistic Model (DDPM) ~\cite{DenoisingDiff2020} in 2020, has demonstrated outstanding performance in producing visually realistic and diverse samples with strong semantic fidelity. Successors such as DDIM~\cite{song2021ddim} and Score-based Generative Models~\cite{song2021scorebased} further accelerated sampling and extended generation to continuous domains.}

\textcolor{black}{Compared to GANs, diffusion models are more stable during training and less prone to mode collapse, making them increasingly favored in high-resolution pathology synthesis. In computational pathology, diffusion models have been applied to gigapixel-scale WSI generation~\cite{harb2023diffusionbasedgenerationhistopathologicalslide}, rare-class augmentation~\cite{oh2023diffmix}, and denoising or artifact removal~\cite{xu2024histodiffusion, 10.1007/978-3-031-66535-6_9}. Recent models have further enabled synthesis conditioned on disease labels~\cite{10635191}, segmentation masks~\cite{shrivastava2023nasdm}, and genomic signals~\cite{carrilloperez2025generation}. To enhance controllability, methods such as Classifier-Free Guidance (CFG, 2022)~\cite{ho2022classifierfreeguidance}, Stable Diffusion~\cite{rombach2022stablediffusion}, and ControlNet~\cite{zhang2023controlnet} were introduced. These allow models to accept user-defined constraints in the form of class labels, text prompts, edge maps, or spatial masks. In pathology, such control mechanisms have been used to synthesize mitosis stages~\cite{bahadir2023characterizing}, generate tumor tiles guided by transcriptomic profiles~\cite{carrilloperez2025generation}, and support counterfactual image generation for model interpretability~\cite{Zigutyte2025}.}

\subsection{LLM and VLM-based Models}
\textcolor{black}{Large language models and vision-language models represent a paradigm shift from pixel-level synthesis to knowledge-based generation. These models leverage large-scale autoregressive Transformers as their core engine, enabling complex multimodal reasoning and the synthesis of new, structured text sequences, such as diagnostic reports and answers to clinical queries. A generative vision–language model contains a decoding component capable of synthesizing novel outputs in at least one modality (vision or language) \cite{genvlm1,genvlm2}. For image-conditioned text generation, this involves a language decoder that generates sequences via autoregressive token prediction or masked-token infilling. Conversely, text-to-image generation relies on an image decoder, such as a diffusion model, to synthesize images from a textual prompt. It is crucial to distinguish these generative VLMs from a parallel lineage of non-generative, contrastive learning models \cite{javed2024cplip}. While the latter have been pivotal in learning powerful} \textcolor{black}{ multimodal representations \cite{ahmed2024pathalign}, they lack the decoder architecture necessary for synthesizing new content and thus fall outside the scope of this part.}

\textcolor{black}{The evolution of generative VLMs was driven by breakthroughs in general-domain models, beginning with the release of ChatGPT in 2022 and the open-source development of multimodal architectures such as LLaVA \cite{llava} in 2023. TraP-VQA \cite{naseem2023vision-language} introduced transformer-based vision–language modeling to pathology, while Med-PaLM M \cite{tu2024towardsgeneralist} extended this paradigm across biomedical data. Domain-adapted models like LLaVA-Med \cite{li2023llavamed} and Quilt-LLaVA advanced specialization for biomedical and pathology-specific VQA. These foundations enabled dedicated architectures such as HistoGPT \cite{Tran2025}, PathCHAT \cite{Lu2024}, and PathologyVLM \cite{Dai2025}, which integrated instruction-tuning and dialogue for report generation and interactive VQA. More recently, LLMs have been used as reasoning engines to guide generative processes, exemplified by TopoFM \cite{li2025topofm}, which employs} \textcolor{black}{topological estimators to synthesize realistic cellular arrangements. This progression underscores the growing role of LLMs as central reasoning components in multi-stage generative pipelines.}

\subsection{Other Generative Approaches}
\textcolor{black}{Beyond the mainstream paradigms, several distinct methodological approaches are emerging. Novel hybrid architectures are being explored, exemplified by models like mSTAR \cite{xu2025multimodalknowledgeenhancedwholeslidepathology} and PathoDuet \cite{hua2024pathoduet}, which employ advanced, multi-stage pre-training strategies to learn powerful, domain-aware representations before the final generative task. Reinforcement learning is also being employed not as a generative architecture itself, but as a powerful fine-tuning strategy to enhance factual accuracy and controllability by optimizing against domain-specific rewards \cite{chen2024starrl}. Looking forward to 2025, the concept of AI Agents marks a leap, positioning optimized generative models as reasoning engines within autonomous systems such as SlideSeek \cite{chen2025evidence} and PathFinder \cite{ghezloo2025pathfinder}, capable of multi-step planning and tool use. These approaches collectively represent a systematic transition toward more modular, controllable, and intelligent generative systems.}
 
\section{Generation Task} 
\label{sec:tasks}
\color{black}
\subsection{Image Generation}
\label{sec:image_generation}
\begin{figure}[!t]
\centerline{\includegraphics[width=0.45\textwidth]{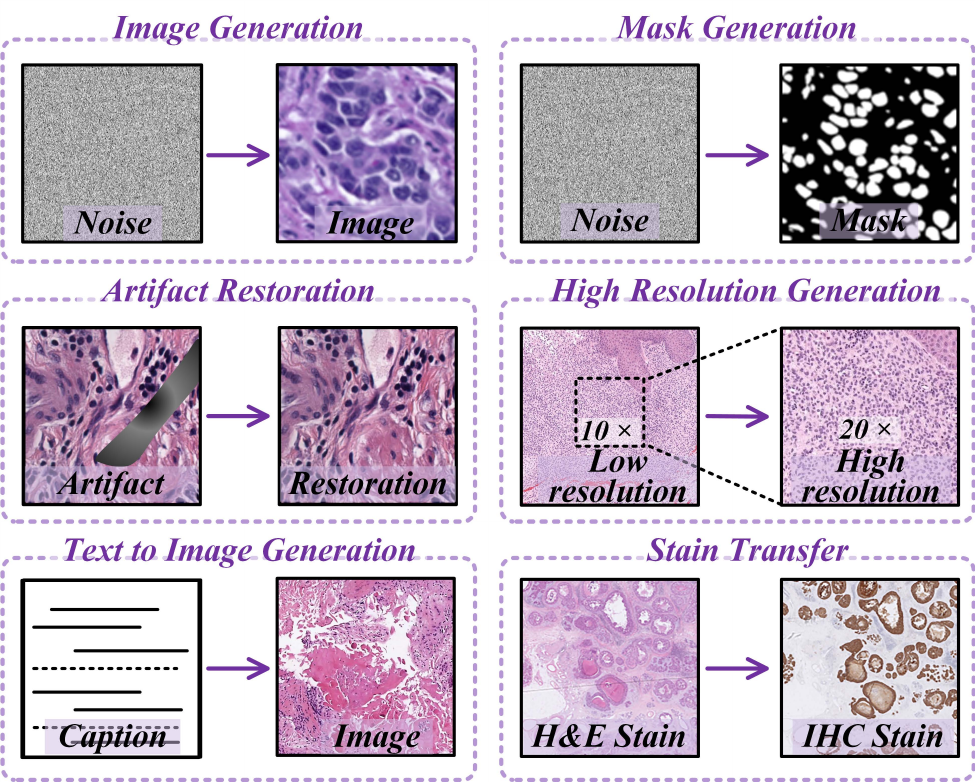}}
\caption{Applications of image generation in computational pathology. Illustrative examples of key tasks, including mask-guided synthesis, resolution enhancement, text-to-image translation, and stain transfer.}
\label{fig_image}
\end{figure}

Image generation plays a foundational role in computational pathology by enabling data augmentation, enhancing visual realism, and preserving structural fidelity, thereby supporting a wide range of downstream tasks (Fig.~\ref{fig_image}). Recent generative models have significantly advanced synthetic pathology imaging, addressing challenges such as limited annotations, class imbalance, and domain shifts. We categorize existing efforts into \textcolor{black}{six} sub-tasks based on their generation goals: (1) synthetic image generation for data augmentation (conditional and unconditional); (2) mask-guided generation for structural fidelity; \textcolor{black}{(3) artifact restoration for removing artifacts and improving image quality;}
(4) high/multiple-resolution generation, including cross-scale images and gigapixel whole-slide generation; (5) text to image generation from diagnostic descriptions or prompts; and (6) stain style generation for normalization and transfer. Each category reflects distinct generative objectives and involves diverse conditioning mechanisms tailored to clinical needs, as illustrated in Tables~\ref{tab:image_generation_part1_fulltext} and \ref{tab:image_generation_part2_fulltext}..

\subsubsection{Synthetic Image and Augmentation}
\label{sec:synthetic_image}
\paragraph{Conditional Image Generation}
Conditional generative models have become the mainstream approach in pathological image synthesis. Their core advantage lies in enabling controllable synthesis guided by semantic, spatial, or biological attributes, thereby providing targeted support for downstream tasks such as segmentation and classification. Early works were predominantly based on GANs, \textcolor{black}{which enabled control over both high-level pathological features like Gleason grade \textcolor{black}{\cite{xue2021selective, golfe2023progleason, li2024unified, vanbooven2025mitigating}} and the fine-grained, realistic composition of individual nuclei for data augmentation \cite{lin2022insmixrealisticgenerativedata}.} Recently, however, the field has witnessed a paradigm shift toward diffusion models, which demonstrate significant advantages in generation quality, training stability, and flexibility of conditioning.

The conditioning mechanisms for diffusion models are evolving towards greater diversity and granularity. A significant body of work focuses on leveraging spatial and semantic priors. These methods aim to enhance the structural accuracy of generated images by introducing various forms of guidance, such as semantic features extracted by Transformers \cite{vitdae2023miccai}, cell nuclei structure maps from two-stage pipelines \cite{Nudiff2024miccai}, custom instance segmentation maps \cite{oh2023diffmix}, or masks provided by unsupervised models like SAM \cite{wang2024usegmix}. While these approaches differ in the source and granularity of their conditional information, they share the common goal of enhancing structural plausibility and image-label alignment, particularly when addressing class imbalance or complex tissue architectures. \textcolor{black}{To bypass costly fine-grained annotations, generation can be conditioned on self-supervised semantic representations, such as unsupervised histological prototypes, which serve as proxies for manual labels \cite{graikos2024learnedrepresentationguideddiffusionmodels, Redekop_2025_CVPR}.}

\color{black}
More advanced research has begun to integrate biological knowledge and weak supervision signals, aiming to improve the clinical interpretability and utility of the models. Multiple studies are moving beyond mere visual realism to simulate authentic biological processes, for instance, by controlling diffusion depth to model the gradual progression of a disease \cite{liu2024ADD}, or using probabilistic mitotic scores as a condition to capture the morphological continuum of cell changes \cite{bahadir2023characterizing}. Furthermore, some research fine-tunes loss weights to emphasize subtle features associated with specific molecular states (e.g., HPV) \cite{porter2024optimising}, while others utilize frameworks like ControlNet with binary images and text prompts to achieve effective control without dense annotations \cite{li2025pdseg, CMPB2023Granddiffusion}. This trend signifies a shift in synthetic image generation from pursuing visually appealing results to creating functionally useful data that reflects underlying biological mechanisms or supports clinical decision-making. Moreover, some work sets the objective of conditional generation directly on functional applications, such as generating counterfactual visual explanations to investigate the impact of biomarkers \cite{Zigutyte2025} or performing anomaly detection via reconstruction error \cite{linmans2024AnoDDPM}, further expanding the application boundaries of generative models.

\paragraph{Unconditional Image Generation}
Unconditional generative models remain indispensable, particularly in unsupervised learning and data-scarce scenarios. Traditional GAN-based methods have played a significant role in learning interpretable latent spaces for phenotype manipulation \cite{quiros2019pathologygan}, self-supervised field-of-view expansion \cite{boyd2021self}, and image restoration \cite{Liang2023}. Notably, some models cleverly bridge the gap between purely unconditional and strongly conditional generation by weakly embedding conditional signals (such as Gleason grades) directly into the generator \cite{golfe2023progleason}. More recently, unconditional diffusion models have been increasingly employed to address specific data challenges.
For example, studies have used them to generate high-quality samples for augmenting training sets, enhancing realism through techniques like color normalization \cite{Moghadam_2023_WACV}; others have focused on generating data for rare modalities like deep ultraviolet (DUV) fluorescence \cite{DUV2024isbi}; and some have mitigated domain shift, \textcolor{black}{for instance by improving segmentation robustness through training-time stain augmentation on synthetically diversified stain styles \cite{bouteldja2022tackling, ktena2024generative}.}
\textcolor{black}{In summary, pathological image synthesis is shifting beyond the pursuit of visual fidelity toward task-oriented and knowledge-driven generation, where models are designed not only to produce realistic images but to serve specific analytical or clinical goals.} \textcolor{black}{Recent work has demonstrated that synthetic images generated by diffusion models can successfully train downstream classifiers to a performance level comparable to those trained on real data \cite{Pozzi2024}. The choice between GANs and diffusion models, and between conditional and unconditional strategies, is now shaped by the demands of downstream tasks, the level of annotation available, and the need for controllability and interpretability in real-world applications.}

\begin{table*}[htbp]
\caption{Overview of image generation methods in computational pathology (Part 1).}
\label{tab:image_generation_part1_fulltext}
\centering
\renewcommand{\arraystretch}{1}
\rowcolors{2}{gray!20}{white}

\begin{tabularx}{\textwidth}{>{\centering\arraybackslash}l l >{\raggedright\arraybackslash}l >{\centering\arraybackslash}l l}
\toprule
\textbf{Year} & \textbf{Method} & \textbf{Input} & \textbf{Arch.} & \textbf{Key Application} \\
\midrule
\multicolumn{5}{l}{\textbf{1. Synthetic Image and Augmentation}} \\
\midrule
\cellcolor{white}2019 & PathologyGAN \cite{quiros2019pathologygan} & Patch & GAN &Deep representations of cancer tissue for classification and generation.\\
\midrule
\cellcolor{white}2020 &Xue et al. \cite{xue2021selective} & Patch & GAN & Dataset augmentation with selective synthetic images for classification. \\
\midrule
\cellcolor{white}&Porter et al. \cite{porter2024optimising} & Patch, HPV Status & Diffusion & Generating images conditioned on HPV status for improved classification. \\
\cellcolor{white}&Bouteldja et al. \cite{bouteldja2022tackling} & Patch & GAN &CycleGAN-based augmentation for stain-invariant segmentation.\\
\cellcolor{white}& insMix \cite{lin2022insmixrealisticgenerativedata} & Patch & GAN & Generating realistic data for augmenting nuclei instance segmentation tasks. \\
\multirow{-4}{*}{2022}\cellcolor{white}&MFDPM \cite{Moghadam_2023_WACV} & Patch & Diffusion & Synthesizing histopathology images preserving morphological details.\\
\midrule
\cellcolor{white} & ViT-DAE \cite{vitdae2023miccai} & Patch & Diffusion & Transformer-driven diffusion autoencoder for augmentation and classification.\\
\cellcolor{white} &Sun et al. \cite{CMPB2023Granddiffusion} & Patch & Diffusion & Instance-aware diffusion model for gland segmentation. \\
\cellcolor{white}& ProGleason-GAN \cite{golfe2023progleason} & Patch & GAN & Synthesizing prostate cancer images with controllable Gleason grades. \\

\cellcolor{white}& DiffMix \cite{oh2023diffmix} & Patch & Diffusion & Synthesizing balanced image-label pairs to improve segmentation. \\
\multirow{-5}{*}{2023}\cellcolor{white}&Bahadir et al. \cite{bahadir2023characterizing} & Patch & Diffusion & Detecting and characterizing mitotic figures for improved classification. \\
\midrule
\cellcolor{white} & AnoDDPM \cite{linmans2024AnoDDPM} & Patch & Diffusion & Detecting out-of-distribution samples in digital pathology images. \\
\cellcolor{white}&Ktena et al. \cite{ktena2024generative} & Patch & Diffusion & Improving medical classifier fairness via generation for detection. \\

\cellcolor{white}&Graikos et al. \cite{graikos2024learnedrepresentationguideddiffusionmodels} & Patch & Diffusion & Representation-guided diffusion models for large-image generation.\\
\cellcolor{white}& ADD \cite{liu2024ADD} & Patch & Diffusion & Simulating progressive pathological transitions for classification tasks. \\

\cellcolor{white}& MoPaDi \cite{Zigutyte2025} & Patch & Diffusion & Generating counterfactual images for mechanistic explanation of AI models. \\
\cellcolor{white}&Li et al. \cite{li2024unified} & Patch & GAN &Unified framework for histopathology image augmentation and classification.\\

\cellcolor{white} &Yu et al. \cite{Nudiff2024miccai} & Patch & Diffusion & Diffusion-based data augmentation for nuclei image segmentation. \\
\cellcolor{white}& USegMix \cite{wang2024usegmix} & Patch & Diffusion & Unsupervised segment mix for data augmentation in classification. \\
\multirow{-9}{*}{2024} \cellcolor{white}&Pozzi et al. \cite{Pozzi2024} & Patch & Diffusion & Evaluating synthetic data quality for downstream classification tasks. \\
\midrule
\cellcolor{white}& PDSeg \cite{li2025pdseg} & Patch, Mask & Diffusion &Patch-wise mask distillation for weakly-supervised tissue segmentation.\\
\multirow{-2}{*}{2025} \cellcolor{white}& \textcolor{black}{dcGAN \cite{vanbooven2025mitigating}} & \textcolor{black}{Patch} & \textcolor{black}{GAN} & \textcolor{black}{Data augmentation with synthetic Gleason patterns to improve Gleason grading.} \\

\midrule
\multicolumn{5}{l}{\textbf{2. Mask-Guided Generation}} \\
\midrule
\cellcolor{white}& AttributeGAN \cite{Attributegan2021miccai} & Cellular Attribute & GAN & Synthesizing histopathology images with control over cellular attributes. \\
\multirow{-2}{*}{2021}\cellcolor{white}& Sharp-GAN \cite{butte2022sharp} & Patch & GAN & Enhancing boundary resolution in mask-to-image synthesis for segmentation. \\
% \midrule

% \multirow{-3}{*}{2022}
\midrule
% & SNOW \cite{ding2023ScientificData} & Patch & GAN & Creating a large-scale synthetic dataset with annotations for nuclei segmentation. \\
\cellcolor{white}&Cechnicka et al. \cite{cechnicka2023realistic} & Patch, Mask & Diffusion & Realistic data enrichment from masks for robust segmentation. \\
\multirow{-2}{*}{2023}\cellcolor{white}& NASDM \cite{shrivastava2023nasdm} & Patch, Mask & Diffusion & Generating nuclei-aware semantic tissue from segmentation masks. \\
\midrule

\cellcolor{white}& DISC \cite{10635191} & Patch & Diffusion & Self-distillation for prostate cancer grading and classification. \\
\cellcolor{white}&Min et al. \cite{Cosynthesis2024eccv} & Text, Point Maps & Diffusion & Co-synthesis of histopathology images and masks from text and point maps. \\
\cellcolor{white}& DiffInfinite \cite{NEURIPS2023_f64927f5} & Patch & Diffusion & Large-scale mask-to-image synthesis using parallel random patch diffusion. \\
\cellcolor{white}& STEDM \cite{ottl2024style} & Patch, Layout & Diffusion & Zero-shot, style-based synthesis of image-mask pairs from semantic layouts. \\
\cellcolor{white}& \textcolor{black}{SynCLay \cite{deshpande2024synclay}} & \textcolor{black}{Layout} & \textcolor{black}{GAN} & \textcolor{black}{Interactive synthesis of images from cellular layouts with co-generated masks.} \\
\multirow{-6}{*}{2024}\cellcolor{white}& HADiff \cite{zhang2025hadiff} & Patch & Diffusion &Hierarchy-aggregated diffusion for mask-based segmentation.\\
\midrule
\cellcolor{white} &Winter et al. \cite{winter2025maskguidedcrossimageattentionzeroshot} & Patch, Mask & Diffusion & Zero-shot, in-silico histopathologic image generation guided by masks. \\
\cellcolor{white}&Jehanzaib et al. \cite{jehanzaib2025robust} & Patch, Mask & GAN & Robust segmentation and data augmentation from masks. \\
\multirow{-3}{*}{2025}\cellcolor{white}& PathoPainter \cite{liu2025pathopainter} & Patch & Diffusion &Tumor-aware image inpainting for segmentation data augmentation.\\

\bottomrule
\end{tabularx}
\end{table*}

\subsubsection{Mask-Guided Generation}
\label{sec:mask-guided}
Mask-guided generation represents a key strategy for enforcing structural awareness in synthetic pathology images, thereby supporting critical applications like data augmentation, segmentation model training, and artifact correction. The primary technical approaches can be categorized by their generation target: direct mask-to-image synthesis and the joint generation of image-mask pairs.

\paragraph{Mask-to-Image Generation}
A major line of work focuses on synthesizing realistic pathology images directly conditioned on segmentation masks. Diffusion models are prevalent in this area due to their high fidelity. Initial methods employed conditional diffusion models to generate images from semantic masks, effectively addressing class imbalance by enriching underrepresented structures \cite{cechnicka2023realistic}. More sophisticated approaches have since refined this process by enhancing the conditioning mechanism. For instance, techniques include using self-distillation from separated conditions for more precise multi-grade synthesis \cite{10635191}, focusing on fine-grained nuclei-level semantics \cite{shrivastava2023nasdm}, or adopting multi-resolution hierarchical features to better preserve the integrity of lesion structures during coarse-to-fine generation \cite{zhang2025hadiff}. In parallel, GAN-based methods offer an alternative pathway. Sharp-GAN, for example, replaces binary masks with normalized nucleus distance maps and introduces a sharpness loss, a design choice that specifically improves boundary resolution for challenging cases like overlapping nuclei \cite{butte2022sharp}.

\paragraph{Joint Image-Mask Generation}
Another research direction aims to simultaneously generate images and their corresponding masks, producing fully paired data ideal for training segmentation models. A key strategy within this domain is the disentanglement of content (the structural mask) from appearance (the visual style). Diffusion models have achieved this by conditioning on both predefined semantic layouts and extracted visual styles \cite{ottl2024style}, or by using cross-image attention to transfer appearance while preserving morphological structures from a reference image \cite{winter2025maskguidedcrossimageattentionzeroshot}. PathoPainter leverages a similar principle for mask-guided inpainting, ensuring strong alignment between the generated image content and the guiding mask \cite{liu2025pathopainter}. To address scalability, hierarchical frameworks like DiffInfinite have been proposed, which first generate a scalable coarse segmentation mask and subsequently synthesize a high-fidelity image via patch-based diffusion conditioned on it \cite{NEURIPS2023_f64927f5}. This architecture enables artifact-free image assembly at arbitrary scales and supports privacy-preserving data augmentation. In contrast, GAN-based approaches provide computationally efficient alternatives with fine-grained control. Methods like Pathopix-GAN are designed for robust, large-scale data augmentation using semantically adaptive normalization \cite{jehanzaib2025robust}, while AttributeGAN demonstrates exceptional control by conditioning generation on specific cellular attributes such as cell crowding and nuclear pleomorphism, enhancing biological interpretability \cite{Attributegan2021miccai}. \textcolor{black}{More recent diffusion models further advance this controllability, enabling the co-synthesis of images and instance labels from intuitive inputs like point maps and text prompts \cite{Cosynthesis2024eccv}, and even user-specified cellular layouts, as demonstrated by frameworks like SynCLay \cite{deshpande2024synclay}.} \textcolor{black}{In summary, mask-guided synthesis enables structural control in pathology image generation. The choice between direct mask-to-image synthesis, joint image–mask generation, and model type (diffusion or GAN) depends on whether the goal is to augment annotations with realistic textures or to create diverse, fully annotated datasets.}

\begin{table*}[htbp]
\caption{Overview of image generation methods in computational pathology (Part 2).}
\label{tab:image_generation_part2_fulltext}
\centering
\renewcommand{\arraystretch}{1}
\rowcolors{2}{gray!20}{white}

\begin{tabularx}{\textwidth}{>{\centering\arraybackslash}l l >{\raggedright\arraybackslash}l >{\centering\arraybackslash}l l}
\toprule
\textbf{Year} & \textbf{Method} & \textbf{Input} & \textbf{Arch.} & \textbf{Key Application} \\
\midrule
\multicolumn{5}{l}{\textbf{3. Artifact Restoration}} \\
\midrule
\multirow{-1}{*}{2020}\cellcolor{white}& Venkatesh et al. \cite{9098358} & Patch & GAN & Restoring marker-occluded regions in histopathology images. \\
\midrule
\multirow{-1}{*}{2022}\cellcolor{white}& MSSA-GAN \cite{Liang2023} & Patch & GAN & Pathology image restoration using multi-scale self-attention. \\
\midrule
\cellcolor{white}& ArtiFusion \cite{he2023artifact} & Patch & Diffusion & Restoring artifacts in histopathology images to improve classification. \\
\cellcolor{white}& AR-CycleGAN \cite{10160043} & Patch & GAN & Detecting and restoring various histological artifacts. \\
\cellcolor{white}& Shen et al. \cite{shen2023federated} & Patch & GAN & Federated learning system for privacy-preserving stain normalization. \\
\multirow{-4}{*}{2023} \cellcolor{white} & Restore-GAN \cite{RONG2023enhanced} & Patch & GAN & Enhancing pathology image quality (deblurring, super-res) for downstream tasks. \\
\midrule
\cellcolor{white}&Wang et al. \cite{10.1007/978-3-031-66535-6_9} & Patch & Diffusion & A lightweight diffusion model for the selective inpainting of histological artifacts. \\
\cellcolor{white}&\textcolor{black}{HARP \cite{pmlr-v250-fuchs24a}} & \textcolor{black}{Patch} & \textcolor{black}{Diffusion} & \textcolor{black}{Artifact restoration through unsupervised detection and diffusion-based inpainting.} \\
\multirow{-3}{*}{2024}\cellcolor{white}& LatentArtiFusion \cite{he2024latentartifusion} & Patch & VAE/Diff. & Restoring histological artifacts efficiently in a low-dimensional latent space. \\
\midrule
\multirow{-1}{*}{2025}\cellcolor{white}& ArtiDiffuser \cite{WANG2025103567} & Patch & Diffusion & Unified framework for histological artifact restoration and synthesis. \\
\midrule
\multicolumn{5}{l}{\textbf{4. High/Multi-Resolution Generation}} \\
\midrule
\cellcolor{white} & ProGAN \cite{levine2020synthesis} & Patch, WSI & GAN & High-quality patch/WSI generation for data augmentation and classification. \\
 \cellcolor{white}&Lahiani et al. \cite{lahiani2021seamless} & Patch & GAN & Synthesizing seamless WSI using perceptual consistency for segmentation. \\
\multirow{-3}{*}{2021}\cellcolor{white}&Boyd et al. \cite{boyd2021self} & Patch & VAE & Generating high-resolution images to improve classification performance. \\
\midrule
\cellcolor{white}\textcolor{black}{2022} & \textcolor{black}{SAFRON \cite{deshpande2022safron}} & \textcolor{black}{Mask} & \textcolor{black}{GAN} & \textcolor{black}{Generating seamless, large-scale histology images via patch-stitching GAN.} \\
\midrule
\cellcolor{white}2023 &Harb et al. \cite{harb2023diffusionbasedgenerationhistopathologicalslide} & WSI & Diffusion & Generating gigapixel-scale whole-slide images from scratch. \\
\midrule
\cellcolor{white} & STAR-RL \cite{chen2024starrl} & Patch & RL &Hierarchical reinforcement learning for interpretable super-resolution.\\
\cellcolor{white}& PathUp \cite{li2024pathup} & Patch, Text & Diffusion & Synthesizing large, multi-class pathology images with high fidelity. \\
\cellcolor{white}& URCDM \cite{cechnicka2024urcdm} & Patch, WSI & Diffusion & Ultra-resolution WSI synthesis using a cascaded conditional diffusion model. \\
\cellcolor{white}&Thakkar et al. \cite{thakkar2024comparative} & Patch & Diffusion & Comparative analysis of diffusion models for synthetic data generation. \\
\multirow{-5}{*}{2024}\cellcolor{white}& Histo-Diffusion \cite{xu2024histodiffusion} & Patch & Diffusion & Super-resolution with quality assessment for classification. \\
\midrule
\cellcolor{white}2025 & ToPoFM \cite{li2025topofm} & Patch & LLM/VLM & Generating images with cellular topology control for segmentation. \\
\midrule
\multicolumn{5}{l}{\textbf{5. Text to Image}} \\
\midrule
2023 \cellcolor{white}& PathLDM \cite{PathLDM2023} & Patch, Text & Diffusion & Generating histopathology images conditioned on descriptive text prompts. \\
\midrule
2024 \cellcolor{white}& VIMs \cite{dubey2024vims_miccai} & Patch, Text & Diffusion & Synthesizing virtual multiplex IHC stains from H\&E images using text prompts. \\
\midrule
\multicolumn{5}{l}{\textbf{6. Stain Synthesis}} \\
\midrule
2017\cellcolor{white} & SST \cite{cho2017neuralstainstyletransferlearning} & Patch & GAN & Stain normalization and style transfer to improve classification robustness. \\
\midrule
2018\cellcolor{white} &Zanjani et al. \cite{zanjani2018stain} & Patch & GAN & Stain normalization using generative adversarial networks. \\
\midrule
\cellcolor{white} & PC-StainGAN \cite{liu2021unpaired} & Patch & GAN & Unpaired stain transfer (e.g., H\&E to IHC) with pathology-consistency constraints. \\
\cellcolor{white} & Residual CycleGAN \cite{DEBEL2021residualcyclegan} & Patch & GAN & Improving domain transformation robustness for segmentation tasks. \\
\multirow{-3}{*}{2021}\cellcolor{white}&Runz et al. \cite{Runz2021normalization} & Patch & GAN & Normalization of H\&E stained images for improved classification. \\
\midrule
\multirow{-1}{*}{2022} \cellcolor{white}& \textcolor{black}{CAGAN \cite{CONG2022102580}} & \textcolor{black}{Patch} & \textcolor{black}{GAN} & \textcolor{black}{Semi-supervised stain normalization via dual-decoder consistency learning.} \\
\midrule
% \cellcolor{white}&Shen et al. \cite{shen2023federated} & Patch & GAN & Federated learning system for privacy-preserving stain normalization. \\
\cellcolor{white} & StainDiff \cite{10.1007/978-3-031-43987-2_53} & Patch & Diffusion & Transferring stain styles between histology images using diffusion models. \\
\multirow{-2}{*}{2023}\cellcolor{white}&Jeong et al. \cite{JEONG2023stain} & Patch & Diffusion & Stain normalization using a score-based diffusion model for classification. \\
\midrule
\cellcolor{white} & AV-GAN \cite{li2024avgan} & Patch & GAN & Virtual staining for unevenly stained tissue using an attention-based varifocal GAN. \\
\cellcolor{white}& StainFuser \cite{jewsbury2024stainfusercontrollingdiffusionfaster} & Patch & Diffusion & Controlling diffusion for fast neural style transfer in multi-gigapixel images. \\
 \cellcolor{white}& StainDiffuser \cite{kataria2024staindiffusermultitaskdualdiffusion} & Patch & Diffusion & Multitask virtual staining and segmentation from the same model. \\
 \cellcolor{white}& PPHM-GAN \cite{Kawai2024Vmia} & Patch & GAN & High-resolution any-to-any stain translation for classification. \\
 \cellcolor{white}& MDDP \cite{lou2024multi-modal} & Patch & Diffusion & Pre-training via H\&E-to-IHC translation for improved WSI classification. \\
 \cellcolor{white}& VirtualMultiplexer \cite{Pati2024} & Patch & Other & AI-based virtual multiplexing for tumor profiling, prognosis, and classification. \\
\cellcolor{white}& TT-SaD \cite{tsai2025test-time} & Patch & Diffusion & Improving classification robustness through test-time stain adaptation. \\
\cellcolor{white}& PST-Diff \cite{he2024pst-diff} & Patch & Diffusion & Achieving high-consistency stain transfer with pathological/structural constraints. \\
\cellcolor{white}& F2FLD \cite{ho2024f2fldmlatentdiffusionmodels} & Patch & Diffusion & Unpaired frozen-FFPE translation for classification. \\
\cellcolor{white}& ULSA \cite{reisenbuchler2024unsupervised} & Patch, WSI & GAN & Unsupervised latent stain adaptation for WSI segmentation. \\
\cellcolor{white}& Diffusion-L \cite{sridhar2024diffusion} & Patch & Diffusion & Image-to-image translation with robust uncertainty quantification. \\
\cellcolor{white}& DUST \cite{Yan2024versatile} & Patch & Diffusion & A unified diffusion framework for versatile (any-to-any) stain transfer. \\
\cellcolor{white}& \textcolor{black}{ODA-GAN \cite{Wang_2025_CVPR}} & \textcolor{black}{Patch} & \textcolor{black}{GAN} & \textcolor{black}{IHC staining via orthogonal decoupling of morphological and staining features.} \\
\multirow{-14}{*}{2024}\cellcolor{white}&Xiong et al. \cite{Xiong_Peng_Zhang_Chen_He_Qin_2025} & Patch & Diffusion & Unpaired multi-domain stain transfer using dual path prompting for classification. \\
\midrule
\cellcolor{white}& CC-WSI-Net \cite{liu2025generatingseamless} & Patch & GAN & Generating seamless virtual IHC whole-slide images with content consistency. \\
\cellcolor{white} & VM-GAN \cite{wang2025value} & Patch & GAN & A value mapping framework for large-scale histological image-to-image translation. \\
\cellcolor{white}& DSTGAN \cite{Du2025deeply} & Patch & GAN & Deeply supervised two-stage GAN for stain normalization and segmentation. \\
\multirow{-4}{*}{2025}\cellcolor{white}& RBDM \cite{zheng2025diffusionbasedvirtualstainingpolarimetric} & Patch & Diffusion & Virtual staining from label-free polarimetric imaging for report generation. \\
\bottomrule
\end{tabularx}
\end{table*}

{\color{black}
\subsubsection{Artifact Restoration}
\label{sec:artifacts_restoration}
Histological artifacts are structural or stylistic distortions introduced during tissue preparation and digitization, such as folds, tears, or staining inconsistencies \cite{taqi2018review}. These artifacts can obscure critical features and impair both human and machine interpretation. Generative models address this issue through two main strategies. (1) Localized restoration focuses on correcting only affected regions while preserving surrounding tissue. GANs have been applied to inpaint occlusions like marker ink \cite{9098358} and reconstruct broader artifact regions with stain and structure preservation \cite{10160043}, with refinements such as multi-scale self-attention enhancing the modeling of contextual dependencies \cite{Liang2023}. Diffusion models adopt masked denoising \cite{he2023artifact}, further enhanced by Transformer-based context modeling \cite{10.1007/978-3-031-66535-6_9} and latent-space optimization \cite{he2024latentartifusion}\textcolor{black}{, and progressing towards fully unsupervised pipelines like HARP \cite{pmlr-v250-fuchs24a} that automate artifact localization prior to restoration.} (2) Holistic enhancement targets global degradations such as blur or inter-site stain variation. Models like Restore-GAN improve overall image quality \cite{RONG2023enhanced}, and federated GANs address distributed inconsistencies \cite{shen2023federated}. Recent advances integrate artifact correction and synthesis into unified frameworks that support both preprocessing and data augmentation, marking a shift toward end-to-end artifact management in computational pathology \cite{WANG2025103567}.}

\subsubsection{High/Multiple-Resolution Generation}
\label{sec:high/multiple-resolution}
The synthesis of diagnostically viable pathology images necessitates reconciling microscopic cellular fidelity with macroscopic anatomical coherence. This fundamental challenge is addressed through two complementary research thrusts: high-resolution generation, which focuses on creating large-scale images with fine spatial detail, and multi-resolution generation, which aims to ensure structural and semantic consistency across different scales of observation.

\paragraph{High-Resolution Generation}
Research in high-resolution synthesis focuses on generating large-scale images with fine spatial detail at a single, high magnification. The predominant strategy to manage the immense pixel space is through progressive or cascaded generation. Early work applied progressive GANs to synthesize 1024$\times$1024 patches \cite{levine2020synthesis} \textcolor{black}{and VAEs to improve classification with high-resolution outputs \cite{boyd2021self}, while parallel efforts focused on ensuring seamless WSI reconstruction by introducing novel losses to eliminate tiling artifacts \cite{lahiani2021seamless}.} This concept has been significantly advanced by diffusion models, which employ coarse-to-fine or cascaded frameworks to sequentially add detail, enabling the synthesis of whole-slide images (WSIs) at the gigapixel scale \cite{harb2023diffusionbasedgenerationhistopathologicalslide, cechnicka2024urcdm}. A parallel effort integrates structural priors, as in ToPoFM, which embeds topology-informed priors into a latent diffusion model to generate high-resolution images with realistic cellular organization, representing a fusion of scalability and structural control \cite{li2025topofm}.

\paragraph{Multi-Resolution Generation}
Multi-resolution image generation in pathology addresses the challenge of ensuring semantic and structural coherence across diagnostic magnifications. Several technical strategies have been explored to achieve this goal. One approach involves explicit alignment and enhancement frameworks. Histo-Diffusion employs a dual-stage pipeline that first restores histological priors and then applies controllable super-resolution, while PathUp incorporates a dedicated “patho-align” module to embed expert priors and mitigate tiling artifacts \cite{xu2024histodiffusion, li2024pathup}. SAFRON \cite{deshpande2022safron} further integrates the stitching process into adversarial training, compelling the generator to produce patches that are seamless and globally consistent.  
A distinct line of work reformulates super-resolution as a reinforcement learning problem. STAR-RL employs a hierarchical agent that adaptively locates low-quality regions and determines termination criteria, enabling efficient and interpretable region-aware reconstruction \cite{chen2024starrl}. Another approach emphasizes scale-aware representation learning. Class-conditioned diffusion models guided by patch size and label prompts explicitly capture relationships between spatial scale and visual features, thereby improving the robustness of downstream classifiers to resolution shifts \cite{thakkar2024comparative}.  Collectively, these strategies highlight complementary pathways toward generating diagnostically consistent images across magnifications. \textcolor{black}{In summary, high-resolution and multi-resolution generation represent complementary dimensions of pathology image synthesis: the former constructs detail-rich gigapixel images, while the latter ensures coherence across diagnostic scales. Clinical applications will require integrated frameworks that preserve both fine detail and cross-scale consistency.}

\subsubsection{Text to Image}
\label{sec:text_to_image}
Text-to-image generation in pathology aims to bridge linguistic clinical descriptions and visual tissue synthesis, enabling controllable and semantically rich image generation. PathLDM \cite{PathLDM2023} introduces a text-conditioned latent diffusion model tailored for histopathology, leveraging GPT-3.5 to summarize diagnostic reports and integrate contextual textual cues into the image generation process. In parallel, VIMs \cite{dubey2024vims_miccai} develops a text-to-stain diffusion framework that synthesizes IHC markers from H\&E-stained (Hematoxylin and Eosin) images using textual prompts, requiring only uniplex IHC training data. Consequently, language-driven generative models hold significant promise for interpretability, modality translation, and report-informed synthesis in digital pathology.
 
\subsubsection{Stain Synthesis}
\label{sec:stain_synthesis}
Generative models help overcome the substantial visual variability in histological images, a key obstacle to developing generalizable AI \cite{breen2024generative}. Stain synthesis comprises two tasks: normalization, which standardizes appearances within a stain type, and transfer, which converts one stain modality into another.

\paragraph{Stain Normalization}
The initial wave of data-driven stain normalization was led by Generative Adversarial Networks (GANs), with unsupervised frameworks like Stain-Style Transfer (SST) network \cite{cho2017neuralstainstyletransferlearning} and CycleGAN \cite{Runz2021normalization} emerging as foundational tools for mapping H\&E images between domains. However, a critical drawback of these early models was the introduction of structural artifacts. This limitation spurred a series of architectural and methodological refinements focused on structure preservation. For instance, Residual CycleGAN introduced a global skip connection to force the model to learn only the residual color transformation, thereby better preserving morphological content \cite{DEBEL2021residualcyclegan}\textcolor{black}{; and frameworks like CAGAN \cite{CONG2022102580} introduced semi-supervised training with dual-decoder consistency to leverage unlabeled source domain data for more robust normalization.} Extending this, DSTGAN integrated a Swin Transformer backbone to better capture long-range contextual information \cite{Du2025deeply}, while other work used InfoGAN-inspired frameworks to explicitly disentangle structure from color attributes \cite{zanjani2018stain}. Beyond image fidelity, research has expanded to address challenges like data privacy, leading to federated cGANs for privacy-preserving normalization across institutions \cite{shen2023federated}. More recently, diffusion models have emerged as a powerful alternative, with score-based methods offering stable normalization \cite{JEONG2023stain}, controlling diffusion for fast, gigapixel-scale style transfer \cite{jewsbury2024stainfusercontrollingdiffusionfaster}, and innovations like TT-SaD enabling robust adaptation without model retraining \cite{tsai2025test-time}\textcolor{black}{, while semi-supervised frameworks like ULSA train models to be inherently stain-invariant by enforcing latent feature consistency on unlabeled data \cite{reisenbuchler2024unsupervised}.}

\paragraph{Stain Transfer}
Representing a more ambitious goal, stain transfer aims to computationally predict one staining modality from another (e.g., H\&E to IHC). Early GAN-based efforts were frequently impaired by tiling artifacts and structural inconsistencies.
To combat this, researchers have introduced pathology-specific consistency constraints \cite{liu2021unpaired}. \textcolor{black}{Advanced models such as ODA-GAN \cite{Wang_2025_CVPR} enforce consistency at the feature level through orthogonal decoupling of staining and morphology, combined with multi-scale biological priors to ensure the clinical utility of virtual IHC panels \cite{Pati2024}.} 
Other approaches employ perceptual losses for improved semantic mapping \cite{lahiani2021seamless}, tiling strategies for whole-slide coherence \cite{liu2025generatingseamless, wang2025value}, \textcolor{black}{and unified frameworks for multi-domain transformations \cite{Kawai2024Vmia}.
Attention-based architectures, such as AV-GAN \cite{li2024avgan}, further address structural inconsistencies in complex regions.}
A definitive paradigm shift is now underway, with diffusion models becoming the dominant technology due to their superior generation fidelity. This has enabled high-fidelity translation from challenging modalities like autofluorescence, label-free polarimetric \textcolor{black}{or artifact-laden frozen sections into virtual stains or their gold-standard counterparts \cite{sridhar2024diffusion, zheng2025diffusionbasedvirtualstainingpolarimetric, ho2024f2fldmlatentdiffusionmodels}.} A key focus in this new paradigm is ensuring structural consistency, leading to training-free frameworks that leverage DDIM inversion \cite{Xiong_Peng_Zhang_Chen_He_Qin_2025} or enforce pathological constraints \cite{he2024pst-diff}. \textcolor{black}{To tackle this from a training perspective, StainDiff \cite{10.1007/978-3-031-43987-2_53} has adapted the cycle-consistency principle, enforcing structural preservation in unpaired stain transfer by applying constraints within the diffusion model's latent space.} The flexibility of diffusion models has also facilitated unified, multi-task architectures capable of any-to-any stain conversion or simultaneous virtual staining and segmentation \cite{Kawai2024Vmia, kataria2024staindiffusermultitaskdualdiffusion, Yan2024versatile}. In a particularly innovative application, virtual staining is now being leveraged as a powerful pretext task for representation learning, where the synthesis objective itself helps a model learn richer semantic features \cite{lou2024multi-modal}. \textcolor{black}{In summary, generative stain synthesis has developed along two complementary paths: normalization, which preserves structural fidelity, and transfer, increasingly enabled by diffusion models with superior realism. Yet clinical deployment depends not only on technical performance; normalization may conceal artifacts and transfer can introduce spurious features, underscoring the necessity of rigorous expert validation.}

\begin{table*}[htbp]
\caption{Overview of text generation methods in computational pathology. }
\label{tab:text_generation_final}
\centering
\renewcommand{\arraystretch}{1}
\rowcolors{2}{gray!20}{white}
\begin{tabularx}{\textwidth}{>{\centering\arraybackslash}l l >{\raggedright\arraybackslash}l >{\centering\arraybackslash}l l}
\toprule
\textbf{Year} & \textbf{Method} & \textbf{Input} & \textbf{Arch.} & \textbf{Key Application} \\
\midrule
\multicolumn{5}{l}{\textbf{1. Image Captioning}} \\
\midrule
2021\cellcolor{white} &Gamper et al. \cite{gamper2021multiple} & Patch & VLM & Generating descriptive captions for pathology patches using a MIL framework. \\
\midrule
2023 \cellcolor{white} & SGMT \cite{qin2023slideimagetellsubtypeguided} & Patch, WSI& VLM & Generating WSI reports guided by cancer information using a masked transformer. \\
\midrule
\cellcolor{white}& HistGen \cite{guo2024histgen} & WSI& VLM & Employing a local-global hierarchical encoder for comprehensive WSI report generation. \\
\multirow{-2}{*}{2024}\cellcolor{white} &Ferber et al. \cite{Ferber2024in-context} & Patch & VLM &In-context learning of large models for pathology image captioning and reporting.\\
\midrule
\multicolumn{5}{l}{\textbf{2. Visual Question Answering (VQA)}} \\
\midrule
\cellcolor{white} & TraP-VQA \cite{naseem2023vision-language} & Patch, Text& VLM & Achieving interpretable VQA by a vision encoder with multimodal attention mechanisms. \\
\multirow{-2}{*}{2023}\cellcolor{white}& LLaVA-Med \cite{li2023llavamed} & Patch, Text& VLM & Building a biomedical VQA assistant through cost-efficient instruction tuning. \\
\midrule
\multirow{-1}{*}{2024} \cellcolor{white}& PathChat \cite{Lu2024} & Patch, Text& VLM & Developing a general-purpose multimodal AI assistant for interactive VQA and reporting. \\
\midrule
\cellcolor{white} & PathologyVLM \cite{Dai2025} & Patch, Text& VLM &A large vision-language model foundational for diverse pathology VQA and reports.\\
\cellcolor{white} & \textcolor{black}{PathCoT \cite{zhou2025pathcotchainofthoughtpromptingzeroshot}} & \textcolor{black}{Patch, Text} & \textcolor{black}{VLM} & \textcolor{black}{Improving reliability of zero-shot CoT reasoning via expert-guided prompting.} \\
\cellcolor{white}& SlideChat \cite{Chen_2025_CVPR} & WSI& VLM & Enabling interactive WSI-level VQA through a hierarchical processing approach. \\
\cellcolor{white}& CLOVER \cite{Chen2025} & Patch, Text& VLM & Cost-effective instruction learning for pathology VQA and object detection tasks. \\
\multirow{-5}{*}{2025}\cellcolor{white} & \textcolor{black}{MUSK \cite{Xiang2025}} & \textcolor{black}{Patch, Text} & \textcolor{black}{VLM} & \textcolor{black}{Leveraging large-scale unpaired data via two-stage pretraining for robust VQA.} \\
\midrule
\multicolumn{5}{l}{\textbf{3. Report Generation}} \\
\midrule

\cellcolor{white}&Sengupta et al. \cite{sengupta2024automaticreportgenerationhistopathology} & WSI& VLM & Achieving end-to-end pathology report generation directly from whole-slide images. \\
\multirow{-3}{*}{2023}\cellcolor{white}& Quilt-LLaVA \cite{Seyfioglu2024quilt-llava} & Patch, WSI& VLM & Enabling spatially-grounded WSI reasoning for VQA and report generation. \\
\midrule
\cellcolor{white} & HistoGPT \cite{Tran2025} & WSI& VLM & Generating human-level, clinical-grade pathology reports from multiple gigapixel WSIs. \\

\cellcolor{white}& MI-Gen \cite{Chen2024wsication} & WSI& VLM & Utilizing a multiple instance generation model for creating detailed WSI reports. \\
% \cellcolor{white}& PathGen-CLIP \cite{sun2024pathgen16m16millionpathology} & Patch & VLM & A CLIP-based approach for simultaneous report generation and classification from patches. \\
\cellcolor{white}& PathGen-LLaVA \cite{sun2024pathgen16m16millionpathology} & Patch & VLM & A visual question answering model trained on PathGen-1.6M dataset. \\
\cellcolor{white}& PathAlign \cite{ahmed2024pathalign} & WSI& VLM & Aligning slide-level visual features with textual reports for improved report generation. \\
\cellcolor{white}& PathInsight \cite{wu2024pathinsightinstructiontuningmultimodal} & Patch, Text& VLM & Instruction-tuning a model on curated datasets for enhanced reporting and VQA. \\
\multirow{-6}{*}{2024}\cellcolor{white}& \textcolor{black}{PRISM \cite{shaikovski2024prismmultimodalgenerativefoundation}} & \textcolor{black}{WSI, Text} & \textcolor{black}{VLM} & \textcolor{black}{A foundation model generating slide-level reports by aggregating tile embeddings.} \\
\midrule
\cellcolor{white} & PolyPath \cite{ahmed2025polypathadaptinglargemultimodal} & WSI, Text& VLM & Adapting a large multimodal model to generate cohesive reports from multiple slides. \\
\cellcolor{white}&Lucassen et al. \cite{lucassen2025pathologyreportgenerationmultimodal} & WSI& VLM & Multimodal representation learning for report generation of melanocytic lesions. \\
\cellcolor{white} & \textcolor{black}{mSTAR \cite{xu2025multimodalknowledgeenhancedwholeslidepathology}} & \textcolor{black}{WSI} & \textcolor{black}{VLM} & \textcolor{black}{Generating reports that directly link tissue morphology to its molecular context.} \\
\cellcolor{white}& KR-MLFS \cite{hu2025pathologyreportgeneration} & WSI& VLM & Integrating knowledge retrieval with multi-level feature selection for report generation. \\
\cellcolor{white}& \textcolor{black}{SlideSeek \cite{chen2025evidence}} & \textcolor{black}{WSI, Text}& \textcolor{black}{VLM} & \textcolor{black}{A multi-agent system that navigates WSIs autonomously for diagnostics.} \\
\cellcolor{white}& \textcolor{black}{TCP-LLaVA \cite{lyu2025efficientslidepathologyvqa}} & \textcolor{black}{WSI, Text} & \textcolor{black}{VLM} & \textcolor{black}{Enabling efficient WSI-level VQA via a novel token compression module.} \\
\cellcolor{white}& \textcolor{black}{Redekop et al. \cite{Redekop_2025_CVPR}} & \textcolor{black}{Prototype} & \textcolor{black}{Diffusion} & \textcolor{black}{Synthesizing images guided by unsupervised prototypes for data-efficient SSL.} \\
\multirow{-8}{*}{2025}\cellcolor{white}& PathFinder \cite{ghezloo2025pathfinder} & WSI, Text& VLM & A multi-agent system that collaborates to generate comprehensive diagnostic reports.\\
\midrule
\multicolumn{5}{l}{\textbf{4. Text-to-Text / Information Extraction}} \\
\midrule
% 2024 \cellcolor{white}& CPLIP \cite{javed2024CLIP} & Text& VLM & Enabling zero-shot retrieval and classification through vision-language alignment. \\
% \midrule
2025 \cellcolor{white}&Saluja et al. \cite{saluja2025cancertypestageprognosis} & Report& LLM & Automated assessment of cancer stage and prognosis from pathology reports.\\
\bottomrule
\end{tabularx}
\end{table*}

\subsection{Text Generation}
\label{sec:text_generation}
\begin{figure}[!t]
\centerline{\includegraphics[width=0.45\textwidth]{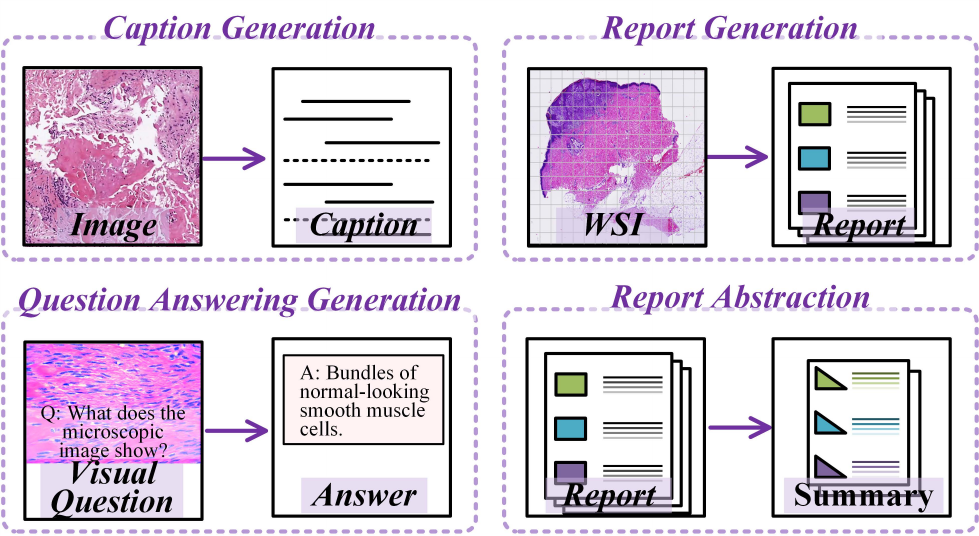}}
% \caption{Illustrative examples of text generation tasks, including text-to-text generation, patch-level image captioning, WSI to report generation, and text synthesis from image-text pairs.
% }
\caption{Applications of text generation in computational pathology. Examples of primary text generation tasks include patch-level captioning, WSI-to-report generation, and text synthesis from image-text pairs.}
\label{fig_text}
\end{figure}

\color{black}
Generative models are fundamentally transforming the interface between visual pathology and clinical language. The application of large language and vision-language models in this domain has evolved along a trajectory of increasing clinical utility. This progression begins with foundational tasks of translating visual features into descriptive text, advances to enabling interactive, dialog-based inquiry of images, \textcolor{black}{through the development of specialized assistants like LLaVA-Med \cite{li2023llavamed},} and culminates in synthesizing comprehensive diagnostic reports. A parallel thrust focuses on leveraging language models for the interpretation and reframing of existing clinical text. Representative examples of these applications are illustrated in Fig.~\ref{fig_text} and Table~\ref{tab:text_generation_final}.

\subsubsection{Image Captioning}
\label{sec:image_captioning}
The foundational task in bridging vision and language is generating accurate textual descriptions for pathological images. Methodological progress has advanced through distinct stages, addressing the unique challenges of gigapixel data and the need for clinical precision. Early work focused on patch-level captioning, adapting standard CNN-encoder and RNN/Transformer-decoder architectures to translate localized visual features into text \cite{tsuneki2022inference, Elbedwehy2024enhanced}. To address the broader context of Whole Slide Images (WSIs), the field shifted towards WSI-level interpretation via multi-instance aggregation. This paradigm treats a WSI as a bag of instances, developing sophisticated mechanisms to synthesize a holistic description. Innovations here include pioneering the use of Multiple Instance Learning (MIL) with dense supervision \cite{gamper2021multiple}, and developing advanced Transformer-based architectures that use techniques like masked attention or query-based aggregation to efficiently link visual information to diagnostic captions \cite{qin2023slideimagetellsubtypeguided, zhou2024pathm3, guo2024histgen}. Most recently, a paradigm shift has been introduced by general-purpose Vision-Language Models (VLMs) like GPT-4V, which leverage in-context learning to perform descriptive tasks with minimal prompting and no domain-specific fine-tuning \cite{Ferber2024in-context}.
\textcolor{black}{In summary, image captioning in pathology has evolved from patch-level descriptions to sophisticated WSI-level interpretation. The field now faces a crossroads between specialized, fine-tuned models offering high precision but requiring extensive data, and flexible, generalist VLMs that raise concerns about reliability and consistency. The central challenge remains validating the clinical and factual accuracy of generated text, regardless of the underlying architecture.}

\subsubsection{Question Answering}
\label{sec:question_answering}
Visual Question Answering (VQA) elevates text generation from description to interactive inquiry, enabling a dynamic dialogue with visual data. \textcolor{black}{A primary challenge in this domain is the scarcity of large-scale, paired image-text data required for robust training. To mitigate this, MUSK \cite{Xiang2025} utilizes a two-stage pretraining strategy that first learns from vast unpaired corpora, while newer systems like PathGen-LLaVA \cite{sun2024pathgen16m16millionpathology} are trained on massive, synthetically generated datasets to overcome this bottleneck. Building upon such data-efficient foundations, another major challenge is scaling VQA from isolated patches to entire WSIs.} \textcolor{black}{This requires addressing both the effective processing of high-resolution patches without information loss and the aggregation of information across the slide. Accordingly, PathologyVLM \cite{Dai2025} introduces a scale-invariant connector to better handle multi-magnification information} \textcolor{black}{inherent in pathology slides, while frameworks like WSI-VQA \cite{chen2025wsi-vqa} and SlideChat \cite{Chen_2025_CVPR} employ hierarchical processing to enable slide-level reasoning. PathChat \cite{Lu2024} serves as a multimodal assistant that integrates a vision encoder with a large language model, enabling pathology-focused question answering and interaction.} A critical component for clinical utility is spatial grounding, the model's ability to localize its answer to specific regions. Models like Quilt-LLaVA have been trained with spatially-aware instruction datasets to achieve this \cite{Seyfioglu2024quilt-llava}. Interpretability is another key focus, with models like TraP-VQA integrating multimodal attention and visualization techniques to explain their reasoning \cite{naseem2023vision-language, he2020pathvqa30000questionsmedical}, \textcolor{black}{prompting strategies like PathCoT \cite{zhou2025pathcotchainofthoughtpromptingzeroshot} structure Chain-of-Thought to ensure its reliability.} Furthermore, to address high computational costs, cost-effective instruction tuning methods leverage powerful LLMs to generate training data for smaller, more efficient models \cite{Chen2025}, \textcolor{black}{while architectural innovations like the token compression in TCP-LLaVA \cite{lyu2025efficientslidepathologyvqa} directly tackle the input sequence length bottleneck.}
\textcolor{black}{In summary, pathology VQA is rapidly maturing from patch-level queries to interactive, slide-level dialogues with spatial awareness. While progress in model architecture and interpretability is significant, the field is constrained by data scarcity and the challenge of clinical validation. The ultimate success of VQA systems will depend on their ability to provide not just answers, but reliably accurate and interpretable answers that can be trusted in a diagnostic setting.}

\subsubsection{Report Generation}
\label{sec:report_generation}
Automated report generation from visual evidence represents a pinnacle task, aiming to synthesize comprehensive, clinical-grade narratives. A key technical challenge is the effective fusion of global WSI context with salient local features. \textcolor{black}{Foundational architectures address this aggregation challenge through methods like compressing tile embeddings with a perceiver network, as seen in PRISM \cite{shaikovski2024prismmultimodalgenerativefoundation}, or by directly constructing multi-scale representations using hierarchical Vision Transformers (HIPT) \cite{sengupta2024automaticreportgenerationhistopathology}.} \textcolor{black}{HistoGPT \cite{Tran2025} successfully scales report generation to the whole-slide level, processing multiple gigapixel WSIs to produce narratives of a quality comparable to human experts and exhibiting emergent zero-shot capabilities. In a conceptual leap beyond this, mSTAR \cite{xu2025multimodalknowledgeenhancedwholeslidepathology} enriched its pretraining by incorporating gene expression data alongside images and reports, enabling the generation of narratives that directly link tissue morphology to its molecular context.} Other approaches focus on enhancing spatial perception through hierarchical position-aware modules in the encoder \cite{Chen2024wsication} or augmenting the generation process with knowledge retrieval decoders \cite{hu2025pathologyreportgeneration}. A significant trend is the move towards emulating the pathologist's holistic workflow. This is exemplified by multi-agent AI frameworks like PathFinder, which uses collaborative agents to triage, navigate, and describe findings, thereby producing an interpretable diagnostic narrative \cite{ghezloo2025pathfinder}. Furthermore, models are being increasingly fine-tuned for specific clinical contexts, such as generating reports for melanocytic lesions or using parameter-efficient techniques like LoRA to adapt large VLMs for creating detailed, part-level reports from multiple slides \cite{lucassen2025pathologyreportgenerationmultimodal, ahmed2024pathalign, ahmed2025polypathadaptinglargemultimodal, wu2024pathinsightinstructiontuningmultimodal}.
\textcolor{black}{In summary, automated report generation has advanced from basic image-to-text mapping toward more complex diagnostic reasoning, integrating multi-scale feature aggregation and cross-slide contextual understanding. Despite notable gains in linguistic fluency and factual accuracy, clinical deployment remains hindered by the critical gap between generating plausible narratives and ensuring diagnostic validity. Achieving clinically error-free outputs demands rigorous, large-scale validation against expert standards.}

\subsubsection{Report Abstraction}
\label{sec:report_abstraction}
Report abstraction has recently emerged as an important application of LLMs in computational pathology. Beyond generating reports from visual inputs, LLMs are now applied to transform unstructured pathology narratives into structured or semantically simplified representations. This enables two main functions: automated extraction of structured data for downstream analysis and decision support, and the generation of accessible summaries to improve communication with non-expert audiences.

For structured information extraction, foundation models such as GPT-4o and Llama 3 have demonstrated annotation-level accuracy in parsing free-text reports into machine-readable forms \cite{Shahid2025usinggenerativeai, balasubramanian2025leveraginglargelanguagemodels}. Instruction-tuned variants, including Path-llama3.1-8B and Path-GPT-4o-mini-FT, further exhibit robust generalization in zero-shot settings for extracting clinically relevant parameters such as cancer type, AJCC stage, and prognosis \cite{saluja2025cancertypestageprognosis}. Complementarily, LLMs have shown promise in patient-facing applications by rephrasing technical content into more interpretable language. Studies using GPT-4 indicate that such interpretive summaries can improve patient comprehension and engagement without compromising diagnostic fidelity \cite{Yang2025enhancing}, thus facilitating informed decision-making and health literacy. \textcolor{black}{In summary, LLM-driven report abstraction enables both clinical efficiency and patient-centered communication. Broader adoption, however, requires rigorous validation across diverse diseases and institutions, with the critical challenge of preserving diagnostic nuance while simplifying language to balance accessibility and fidelity.}

\begin{table*}[!t]
\caption{Overview of generative methods bridging histology and genomics for cross-modal prediction and generation. \\ \scriptsize Abbreviations: ST, Spatial Transcriptomics; scRNA-seq, Single-cell RNA sequencing; TME, Tumor Microenvironment.}
\label{tab:gene_generation_simplified}
\centering
\renewcommand{\arraystretch}{1}
\rowcolors{2}{gray!20}{white}
\begin{tabularx}{\textwidth}{>{\centering\arraybackslash}l >{\raggedright\arraybackslash}l >{\raggedright\arraybackslash}p{3.5cm}>{\centering\arraybackslash}l >{\raggedright\arraybackslash}X}
\toprule
\textbf{Year} & \textbf{Method} & \textbf{Input} & \textbf{Arch.} & \textbf{Key Application} \\
\midrule
\multicolumn{5}{l}{\textbf{1. Virtual Molecular Profiling}} \\
\midrule
2023 \cellcolor{white}& SpatialScope \cite{wan2023integrating} & ST, scRNA-seq & VAE & Deconvoluting and imputing ST data to achieve single-cell, whole-transcriptome resolution. \\
\midrule
2024\cellcolor{white} & Diff-ST \cite{wang2024cross-modal} & Low-Resolution ST, Patch & Diffusion & Super-resolving spatial transcriptomics using histology as guidance. \\
\cellcolor{white} & SCHAF \cite{comiter2024inference} & Patch, ST & VAE & Cross-modal prediction of single-cell gene expression from histology.\\
\midrule
\cellcolor{white} & PAST \cite{yang2025pastmultimodalsinglecellfoundation} & Patch, scRNA-seq & / & Single-cell gene expression and virtual IHC prediction from cell images. \\
\cellcolor{white} & SPATIA \cite{kong2025spatiamultimodalmodelprediction} & Patch, scRNA-seq & / & Cross-modal generation between Patch and scRNA-seq (Bidirectional) \\
\cellcolor{white} & GenST \cite{wood2025genst} & Patch, ST & VAE & Predicting spatial transcriptomics by aligning cross-modal latent spaces. \\
% \cellcolor{white} & SCHAF \cite{comiter2024inference} & Patch, ST & VAE & Cross-modal prediction of single-cell gene expression from histology.\\
\cellcolor{white}& \textcolor{black}{Stem \cite{zhu2025diffusiongenerativemodelingspatially}} & \textcolor{black}{Patch}& \textcolor{black}{Diffusion} &\textcolor{black}{Spatial transcriptomics inference via conditional diffusion models.}\\
\cellcolor{white}& \textcolor{black}{LD-CVAE \cite{Zhou2025RobustMS}} & \textcolor{black}{WSI}& \textcolor{black}{VAE} & \textcolor{black}{Predicting surrogate genomic embeddings from WSI for survival prediction.} \\
\multirow{-7}{*}{2025}\cellcolor{white} & HistoPlexer \cite{andani2025histopathology} & Patch, Protein Multiplex & GAN & Generating multiplex protein maps from H\&E for TME characterization. \\
\midrule
\multicolumn{5}{l}{\textbf{2. Reverse Morphology Generation }} \\
\midrule
2024 \cellcolor{white}& Pix2Path \cite{fu2024pix2path} & High-Resolution ST, Patch & GAN & Generating pathology images from ST for risk and perturbation analysis. \\
\midrule

\cellcolor{white} & RNA-CDM \cite{carrilloperez2025generation} & Bulk RNA-seq & Diffusion & Generating pathology images from bulk RNA-seq for augmentation. \\
\multirow{-1}{*}{2025}\cellcolor{white} & HistoXGAN \cite{howard2024generative} & Pathologic, genomic, and radiographic Feature  & GAN & Reconstructing histology from latent multi-modal features for explainability and virtual biopsy. \\
\bottomrule
\end{tabularx}
\end{table*}  
 
\subsection{Molecular Profiles-Morphology Generation}
\label{sec:molecular_profiles}
\textcolor{black}{The fundamental challenge of linking genotype to phenotype in biomedicine manifests acutely in oncology, where molecular alterations must be reconciled with observable morphological changes. Traditional approaches compartmentalize these domains: histopathological analysis extracts spatial architectural information from H\&E-stained sections, while omics profiling quantifies molecular features at the expense of spatial context \cite{staahl2016visualization}. Spatial omics technologies, though promising, remain constrained by cost and technical complexity, limiting their clinical penetration \cite{rao2021exploring}.
Generative models learn the joint distribution of histology and molecular profiles, which enables conditional inference in two complementary directions in Table~\ref{tab:gene_generation_simplified}: inferring molecular profiles from histology, termed virtual molecular profiling, and synthesizing histology from molecular inputs, termed reverse morphology generation.}

\subsubsection{Virtual Molecular Profiling}
\label{sec:virtual_molecular_profiling}
\textcolor{black}{Virtual molecular profiling aims to transform large repositories of low-cost H\&E slides into computable molecular maps, offering transcriptomic or proteomic readouts without additional assays.  A primary objective is the de novo synthesis of spatial transcriptomes. Large-scale foundation models like PAST achieve this at single-cell resolution \cite{yang2025pastmultimodalsinglecellfoundation}, while alternative frameworks such as SCHAF employ graph-based architectures to incorporate structural priors for the same task \cite{comiter2024inference}. Lightweight models like GenST offer efficient prediction through aligned autoencoders \cite{wood2025genst}. \textcolor{black}{To better capture biological heterogeneity, Stem \cite{zhu2025diffusiongenerativemodelingspatially} reframes deterministic regression as conditional diffusion-based generation, yielding distributions of plausible expression profiles from a single H\&E patch.} This generative capability extends to proteomics, where HistoPlexer synthesizes multiplexed protein maps while preserving biologically crucial co-localization patterns \cite{andani2025histopathology}. Beyond de novo synthesis, other models focus on enhancing existing molecular data. For instance, Diff-ST uses histology as a morphological prior to super-resolve low-resolution spatial transcriptomics \cite{wang2024cross-modal}, whereas SpatialScope computationally deconstructs mixed-cell data to achieve near single-cell resolution \cite{wan2023integrating}. Generative imputation is critical for clinical robustness, as models like LD-CVAE \cite{Zhou2025RobustMS} synthesize surrogate genomic embeddings from histology to stabilize multimodal predictions when molecular data are absent.}

\subsubsection{Reverse Morphology Generation}
\label{sec:reserve_morphology_generation}
\textcolor{black}{Reverse morphology generation serves as a powerful platform for basic biological discovery and model interpretability. By generating histology images conditioned on molecular profiles, these models provide visual hypotheses for how molecular states shape tissue architecture, creating an effective in silico experimental system. This task can be driven by inputs with varying levels of spatial information. Models like Pix2Path synthesize pathology images from high-resolution spatial transcriptomics, a capability that can be leveraged for virtual gene perturbation experiments to probe causal relationships between genes and phenotypes \cite{fu2024pix2path}. Addressing a more formidable challenge, RNA-CDM generates plausible tissue structures from non-spatial bulk RNA-seq data \cite{carrilloperez2025generation}. A distinct application of this reverse-generative principle is for model interpretability. HistoXGAN \cite{howard2024generative} reconstructs histology from latent vectors linked to molecular states, such as PIK3CA mutations, enabling visualization of subtle AI-learned morphological features.}

\textcolor{black}{Despite rapid progress, significant challenges persist, including the scarcity of high-quality, spatially aligned paired datasets \cite{yang2025pastmultimodalsinglecellfoundation, kong2025spatiamultimodalmodelprediction} and the difficulty in validating the biological fidelity of generated outputs. Future progress will depend on building efficient, general-purpose foundation models that support bidirectional inference, exemplified by frameworks like SPATIA \cite{kong2025spatiamultimodalmodelprediction}. The ultimate objective is to create a digital twin of tissues, a robust computational model capable of accurately simulating the morphological consequences of genetic interventions and drug responses.}

\begin{table*}[!t]
\caption{Overview of other specialized generative methods in computational pathology.}
\label{tab:other}
\centering
\renewcommand{\arraystretch}{1}
\rowcolors{2}{gray!20}{white}
\begin{tabularx}{\textwidth}{>{\centering\arraybackslash}l l >{\raggedright\arraybackslash}l >{\centering\arraybackslash}l l}
\toprule
\textbf{Year} & \textbf{Method} & \textbf{Input} & \textbf{Arch.} & \textbf{Key Application} \\
\midrule
\multicolumn{5}{l}{\textbf{1. Spatial Layout Generation}} \\
\midrule
\cellcolor{white}&Li et al. \cite{Diffusioncelllayout2024miccai} & Patch, Layout& Diffusion & Guiding image generation and object detection using explicit spatial layout priors. \\
\multirow{-2}{*}{2024}\cellcolor{white}& TopoCellGen \cite{TopoCellGen2025CVPR} & Cell Layout& Diffusion & Generating biologically plausible, topology-aware cell layouts for downstream tasks. \\
\midrule
2025 \cellcolor{white}& DAMM-Diffusion \cite{zhou2025dammdiffusionlearningdivergenceawaremultimodal} & Patch& Diffusion & Predicting nanoparticle distributions within tumor microenvironments for prognosis. \\
\midrule
\multicolumn{5}{l}{\textbf{2. Semantic Output Generation}} \\
\midrule
 \cellcolor{white}& QAP \cite{yin2024QAP} & Patch, WSI& VLM &Quantitative visual prompts from histological features for model adaptation. \\
\multirow{-2}{*}{2024}\cellcolor{white}& TQx \cite{nguyen2024towardsatext-based} & Patch& VLM & Generating text-based image embeddings for explainable classification and clustering. \\
\midrule
2025 \cellcolor{white}& MLLM4PUE \cite{zhou2025mllm4pueuniversalembeddingsdigital} & Patch, Text& VLM & Generating universal embeddings for zero-shot classification and retrieval tasks. \\
\midrule
\multicolumn{5}{l}{\textbf{3. Latent Representation Generation}} \\
\midrule
2019 \cellcolor{white}&Hu et al. \cite{8402089} & Patch& GAN &Unsupervised visual representations from patches for segmentation and classification. \\
\midrule
2023 \cellcolor{white}& PLIP \cite{Huang2023} & Patch, Text& VLM &Learning joint image-text embeddings from web data for retrieval and classification. \\
\midrule
\cellcolor{white}& PRDL \cite{Tang2025promptable} & Patch, WSI& MIL &Prompt-guided sampling for diverse WSI representations in MIL pipelines. \\
\cellcolor{white}& AugDiff \cite{10666706} & Patch, WSI& Diffusion &Generating semantic feature augmentations to improve classification generalization. \\
\cellcolor{white}& DCDiff \cite{10577168} & Patch, WSI& Diffusion & Generating multi-resolution features for robust classification. \\
\cellcolor{white}& \textcolor{black}{Prov-GigaPath \cite{Xu2024}} & \textcolor{black}{WSI} & \textcolor{black}{/} & \textcolor{black}{Generating slide-level features for downstream tasks.} \\
\multirow{-5}{*}{\cellcolor{white}2024}& \textcolor{black}{GPFM \cite{ma2025generalizablepathologyfoundationmodel}} & \textcolor{black}{Patch} & \textcolor{black}{/} & \textcolor{black}{Generating universal feature backbone via multi-expert knowledge distillation.} \\
\midrule
2025 \cellcolor{white}& MExD \cite{nguyen2025mgpathvisionlanguagemodelmultigranular} & Feature& Diffusion &Mixture-of-experts class distribution synthesis for generative WSI classification. \\
\midrule
\multicolumn{5}{l}{\textbf{4. Cell Simulation}} \\
\midrule
 \cellcolor{white}& SynCellFactory \cite{SynCellFactory2024miccai} & Patch& Diffusion &Synthesizing realistic time-lapse cell videos with ground truth for tracking tasks. \\
\multirow{-2}{*}{2024}\cellcolor{white}&Bruch et al. \cite{bruch2024improving3ddeeplearning} & Patch & GAN &Coherent 3D cellular structures and masks by incorporating biophysical constraints. \\
\bottomrule
\end{tabularx}
\end{table*}

\subsection{Other Generation}
\label{sec:other_generation}
Recent advances in generative modeling for pathology have expanded beyond traditional image and text synthesis to target more specialized data modalities and objectives. Current efforts can be grouped into four emerging directions, as summarized in Table~\ref{tab:other}. The first focuses on spatial layout generation, simulating biologically realistic tissue and cellular organizations under structural constraints. The second emphasizes semantic output generation, producing interpretable artifacts such as prompts or embeddings to enhance model transparency and human–AI interaction. The third centers on latent representation generation, shifting synthesis from the pixel space to the latent space to improve efficiency, scalability, and downstream performance. Finally, cell simulation applies generative models to synthesize photorealistic microscopy data with programmatically defined ground truth, alleviating annotation bottlenecks for tasks such as tracking and segmentation.

\subsubsection{Spatial Layout Generation}
\label{sec:spacial_layout}
This category of generative models aims to simulate biologically realistic spatial configurations of cellular or tissue structures, enabling downstream analysis under controlled topological assumptions. The core challenge is to preserve complex inter- and intra-cellular spatial relationships. To achieve this, recent methods impose strong topological or spatial constraints on the generation process. For example, TopoCellGen integrates persistent homology into a diffusion framework to maintain spatial fidelity, while other diffusion-based approaches are guided by density-based cell layout maps \cite{TopoCellGen2025CVPR, Diffusioncelllayout2024miccai}. An alternative models the cellular environment as a graph, with DiGress \cite{DiGress2024miccai} employing graph-based diffusion to generate cell graphs that preserve structural characteristics such as tertiary lymphoid structures. These techniques further generalize to multimodal settings, exemplified by DAMM-Diffusion \cite{zhou2025dammdiffusionlearningdivergenceawaremultimodal}, which predicts nanoparticle distributions by fusing structural information from the tumor microenvironment.
\textcolor{black}{In summary, generative spatial layout models have progressed from simple point patterns to structured tissue ecosystems, offering realistic synthetic microenvironments for biology and analysis, though validating their fidelity to complex biological rules remains a key challenge.}

\subsubsection{Semantic Output Generation}
\label{sec:semantic_output}
This area seeks to produce interpretable artifacts, including prompts and embeddings, that improve model transparency and facilitate human–AI interaction. One research direction translates histological features into structured visual prompts for downstream guidance, as exemplified by QAP \cite{yin2024QAP}. A second line of work develops multimodal large language models that derive universal embeddings from summarization-style prompts, enabling robust zero-shot classification and retrieval \cite{zhou2025mllm4pueuniversalembeddingsdigital}. A third direction aligns visual features with semantically rich spaces through training on large-scale, internet-curated image–text datasets, as in PLIP \cite{Huang2023}. Frameworks such as TQx \cite{nguyen2024towardsatext-based} extend this approach by generating text-based image embeddings constructed from domain-specific vocabularies, yielding representations that are inherently explainable. \textcolor{black}{In summary, semantic output generation provides a bridge between computational models and clinical workflows by producing interpretable, knowledge-aligned outputs, though a key challenge remains to ensure that these semantic concepts are consistently grounded in true pathological features.}

\subsubsection{Latent Representation Generation}
\label{sec:latent_representation}
This paradigm reorients generation from pixel-level synthesis to the more abstract and efficient latent space, where intermediate feature representations are synthesized or augmented to enhance downstream robustness and performance. For instance, diffusion models can be applied directly to latent features to generate semantically coherent augmentations, a technique that has proven highly effective for improving WSI classification while avoiding costly image transformations \cite{10666706}. Building on this idea, PRDL employs prompt-guided sampling in the latent space to create diverse WSI representations for multiple instance learning \cite{Tang2025promptable}. \textcolor{black}{Recent advances have extended this paradigm to gigapixel WSIs, as demonstrated by Prov-GigaPath \cite{Xu2024}, which adapts the LongNet transformer to learn holistic latent representations from tiles, establishing a benchmark for slide-level feature generation.} MExD \cite{nguyen2025mgpathvisionlanguagemodelmultigranular} combines a mixture-of-experts aggregator with a diffusion-based generative classifier to synthesize class distributions for WSI classification, while DCDiff \cite{10577168} employs a dual-granularity diffusion architecture to generate features at multiple spatial scales. \textcolor{black}{Another line of work distills knowledge from multiple expert models to construct universal latent representations, exemplified by the Generalizable Pathology Foundation Model (GPFM) \cite{ma2025generalizablepathologyfoundationmodel}. In summary, latent representation generation offers an efficient alternative to pixel-level synthesis, significantly enhancing downstream performance, although concerns remain regarding the biological interpretability of the resulting features.}

\begin{figure*}[!t]
\centerline{\includegraphics[width=\textwidth]{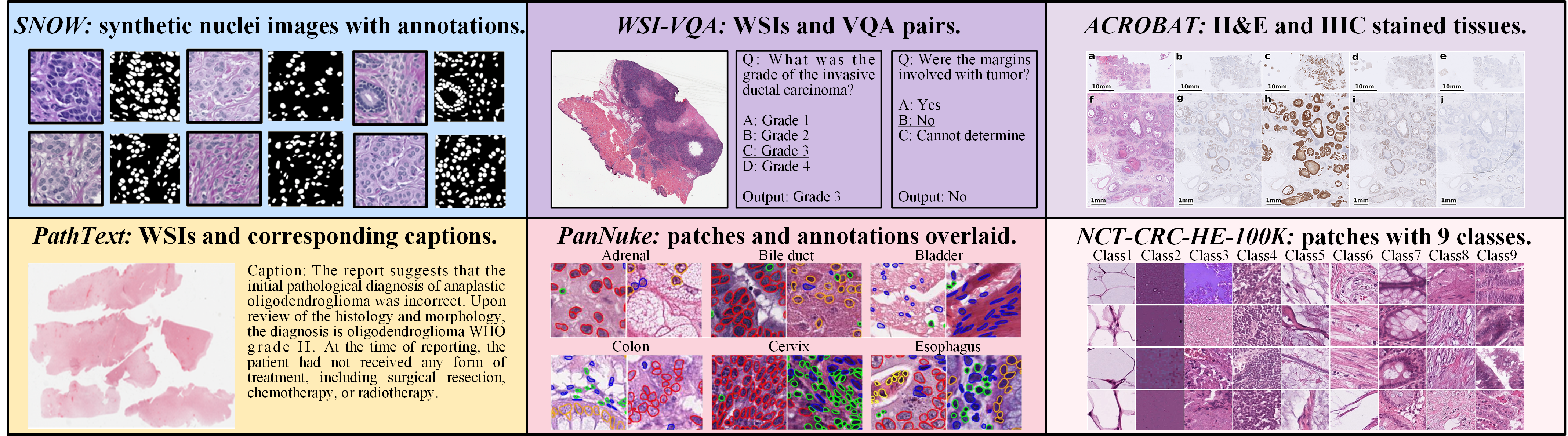}}
\caption{Examples from representative pathology datasets showcasing variations in tissue type, staining, annotations, captions, and VQA pairs.}
\label{fig_dataset}
\end{figure*}

\begin{table*}[!t]
\caption{Datasets for content generation models in computational pathology, organized by task and data type. Pairs indicate image–text pairs.}
\label{tab:dataset_optimized}
\centering
\renewcommand{\arraystretch}{1}
\rowcolors{2}{gray!20}{white}
\resizebox{\textwidth}{!}{
\begin{tabular}{@{}lp{1.2cm}p{1.4cm}p{1.5cm}p{2.6cm}p{8.5cm}@{}}
\toprule
\textbf{Dataset} & \textbf{Organ} & \textbf{Modality} & \textbf{Stain Type} & \textbf{Size} & \textbf{Source} \\
\midrule
\multicolumn{6}{c}{\textbf{I. Real-World Benchmark Datasets}} \\
\midrule
\multicolumn{6}{@{}l}{\textbf{A. Segmentation \& Detection}} \\
Camelyon16 \cite{10.1001/jama.2017.14585} & Breast & WSI & H\&E & 400 WSIs & \url{https://camelyon16.grand-challenge.org} \\
CoNIC \cite{graham2023conicchallengepushingfrontiers} & Colon & Patch & H\&E & 4,981 images & \url{https://conic-challenge.grand-challenge.org} \\
CoNSeP \cite{graham2019hover} & Colon & Patch & H\&E & 41 images & \url{https://opendatalab.com/OpenDataLab/CoNSeP} \\
DigestPath \cite{DA2022102485} & Multiple & Patch & H\&E & 1,559 images & \url{https://digestpath2019.grand-challenge.org} \\
GlaS \cite{sirinukunwattana2016glandsegmentationcolonhistology} & Multiple & Patch & H\&E & 165 images & \url{https://www.kaggle.com/datasets/sani84/glasmiccai2015-gland-segmentation} \\
Lizard \cite{graham2021lizard} & Colon & Patch & H\&E & 495,179 nuclei & \url{https://www.kaggle.com/datasets/aadimator/lizard-dataset} \\
MoNuSeg \cite{8880654} & Multiple & Patch & H\&E & 29,000 nuclei & \url{https://monuseg.grand-challenge.org/} \\
PanNuke \cite{gamper2019pannuke} & Multiple & Patch & H\&E & 7,901 images & \url{https://huggingface.co/datasets/RationAI/PanNuke} \\
TUPAC16 \cite{VETA2019111} & Breast & WSI & H\&E & 573 cases & \url{https://tupac.grand-challenge.org} \\
\midrule
\multicolumn{6}{@{}l}{\textbf{B. Classification}} \\
BACH \cite{polonia_2019_3632035} & Breast & Patch, WSI & H\&E & 400+ images & \url{https://zenodo.org/records/3632035} \\
BreakHis \cite{7312934} & Breast & Patch & H\&E & 9,709 images & \url{https://web.inf.ufpr.br/vri/databases/breakhis} \\
Camelyon17 \cite{8447230} & Breast & WSI & H\&E & 1,399 WSIs & \url{https://camelyon17.grand-challenge.org/} \\
LC25000 \cite{andrewlc25000} & Multiple & Patch & H\&E & 25,000 images & \url{https://huggingface.co/datasets/1aurent/LC25000} \\
NCT-CRC-HE-100K \cite{kather_2018_1214456} & Colon & Patch & H\&E & 100,000 images & \url{https://zenodo.org/records/1214456} \\
PANDA \cite{Bulten2022} & Prostate & WSI & H\&E & 11,000 WSIs & \url{https://panda.grand-challenge.org} \\
PatchCamelyon \cite{b_s_veeling_j_linmans_j_winkens_t_cohen_2018_2546921} & Multiple & Patch & H\&E & 327,680 images & \url{https://github.com/basveeling/pcam} \\
TCGA-BRCA & Multiple & WSI & H\&E & 1,098 cases & \url{https://portal.gdc.cancer.gov/projects/TCGA-BRCA} \\
\midrule
\multicolumn{6}{@{}l}{\textbf{C. Stain Transfer}} \\
ACROBAT \cite{weitz2022acrobatmultistainbreast} & Breast & WSI & H\&E, IHC & 4,212 WSIs & \url{https://acrobat.grand-challenge.org} \\
BCI \cite{Liu_2022_CVPR} & Breast & WSI & H\&E, IHC & 9746 images & \url{https://bci.grand-challenge.org} \\
EMPaCT \cite{karkampouna_2023_10066853} & Prostate & Patch & H\&E, IHC & 420 images & \url{https://zenodo.org/records/10066853} \\
\midrule
\multicolumn{6}{@{}l}{\textbf{D. Multimodal \& Other Tasks (Retrieval, Captioning, Biomarker Exploration)}} \\
ARCH \cite{gamper2020multiple} & Multiple & Patch & / & 15,164 images & \url{https://warwick.ac.uk/fac/cross_fac/tia/data/arch} \\
HEST-1k \cite{jaume2024hest} & Multiple & WSI, omics & H\&E & 1,229 cases & \url{https://huggingface.co/datasets/MahmoodLab/HEst} \\
OpenPath \cite{Huang2023} & Multiple & Patch, Text & / & 208,414 images & \url{https://huggingface.co/spaces/vinid/webplip} \\
PathCap \cite{sun2024pathasst} & Multiple & Patch, Text & / & 207K pairs & \url{https://huggingface.co/datasets/jamessyx/PathCap} \\
PatchGastricADC22 \cite{masayuki_tsuneki_2021_6021442} & Stomach & Patch, Text & H\&E & 262K images & \url{https://zenodo.org/records/6021442} \\
QUILT-1M \cite{wisdom_oluchi_ikezogwo_2023_8239942} & Multiple & Patch & / & 1M pairs & \url{https://zenodo.org/records/8239942} \\
\midrule
\multicolumn{6}{c}{\textbf{II. Generative \& Synthetic Datasets}} \\
\midrule
\multicolumn{6}{@{}l}{\textbf{A. Image-Synthetic}} \\
SNOW \cite{ding2023ScientificData} & Breast & Patch & H\&E & 20,000 images & \url{https://zenodo.org/records/6633721} \\
\midrule
\multicolumn{6}{@{}l}{\textbf{B. Text-Synthetic (for VQA \& Report Generation)}} \\
PathGen-1.6M \cite{sun2024pathgen16m16millionpathology} & Multiple & Patch, Text & Multiple & 1.6M  pairs & \url{https://huggingface.co/datasets/jamessyx/PathGen} \\
PathMMU \cite{sun2024pathmmumassivemultimodalexpertlevel} & Multiple & Patch, Text & / & 24K images / 33K QA & \url{https://huggingface.co/datasets/jamessyx/PathMMU} \\
PathText \cite{Chen2024wsication} & Multiple & Patch, Text & / & 9,009 WSI-text pairs & \url{https://github.com/cpystan/Wsi-Caption} \\
PathVQA \cite{he2020pathvqa30000questionsmedical} & Multiple & Patch, Text & / & 5K images / 32K QA & \url{https://huggingface.co/datasets/flaviagiammarino/path-vqa} \\
PMC-VQA \cite{zhang2024pmcvqavisualinstructiontuning} & Multiple & Patch, Text & Multiple & 227K VQA & \url{https://huggingface.co/datasets/RadGenome/PMC-VQA} \\
QUILT-VQA \cite{Seyfioglu2024quilt-llava} & Multiple & Patch, Text & / & 985 images / 1,283 QA & \url{https://huggingface.co/datasets/wisdomik/Quilt_VQA} \\
WSI-VQA \cite{chen2025wsi-vqa} & Breast & WSI, Text & H\&E & 977 WSIs / 8,672 QA & \url{https://github.com/cpystan/WSI-VQA} \\
\bottomrule
\end{tabular}
}
\end{table*}

\subsubsection{Cell Simulation}
\label{sec:cell_simulation}
This highly specialized application of generative models focuses on creating synthetic microscopy data tailored for specific downstream tasks, primarily to address the fundamental challenge of limited annotated data. The goal is to generate photorealistic data that comes with perfect, programmatically generated ground-truth annotations. For instance, to solve the cell tracking annotation bottleneck, SynCellFactory \cite{SynCellFactory2024miccai} uses ControlNets to decouple cellular appearance from motion dynamics, enabling the synthesis of time-lapse videos with corresponding ground-truth trajectories. Similarly, for 3D segmentation, physics-informed GANs can simultaneously generate coherent 3D cellular structures and their corresponding pixel-perfect segmentation masks, using biophysical constraints to ensure the results are biologically plausible \cite{bruch2024improving3ddeeplearning}.
\textcolor{black}{In summary, generative cell simulation offers a targeted means of alleviating annotation bottlenecks in biomedical imaging by producing synthetic datasets with programmatically defined ground truth. Its utility, however, remains constrained by domain shift, since the reliability of downstream models depends on simulations that accurately reflect biological variability and imaging artifacts.}

\section{Dataset}
\label{sec:datasets}
\color{black}
To support the diverse generative tasks in computational pathology, we summarize commonly used datasets spanning a wide range of applications, including classification, segmentation, stain transformation, caption generation, and question answering, as detailed in Table~\ref{tab:dataset_optimized}. \textcolor{black}{Commonly used evaluation metrics are summarized in Supplementary Materials Table I.} These datasets span a broad range of organs, modalities (e.g., whole-slide images, image patches, and textual reports), and staining types (e.g., H\&E and IHC), as illustrated in Fig.~\ref{fig_dataset}, offering a comprehensive foundation for the training and evaluation of generative models in computational pathology. Notably, with the recent development of generative pathology, several synthetic benchmarks have emerged. These include both image-synthetic datasets, such as SNOW \cite{ding2023ScientificData}, which introduces a large-scale dataset of synthetic pathological images paired with semantic segmentation annotations for nuclei; and text-synthetic datasets, where generated textual reports or QA pairs are increasingly used to augment image-text training corpora. Examples include PathGen-1.6M \cite{sun2024pathgen16m16millionpathology}, PathMMU \cite{sun2024pathmmumassivemultimodalexpertlevel}, and WSI-VQA \cite{chen2025wsi-vqa}, which leverage large-scale language models to generate descriptive captions and question-answer pairs.

\section{Discussion}
\label{sec:discussion}
\textcolor{black}{Generative artificial intelligence is increasingly reshaping computational pathology, impacting diagnostic imaging, knowledge extraction, and clinical decision-making. This discussion first examines current capabilities and demonstrated impact, then analyzes critical limitations and barriers, and finally outlines strategic pathways with future directions toward trustworthy clinical adoption.}
\subsection{Current Capabilities and Impact}
\subsubsection{Architectural Suitability and Task Performance}
\textcolor{black}{The optimal choice of a generative model in computational pathology is not absolute but is dictated by a fundamental trade-off among synthesis fidelity, computational efficiency, and architectural inductive bias. The absence of standardized, clinically meaningful benchmarks currently precludes direct numerical comparison across studies, making an understanding of these architectural trade-offs paramount. For image synthesis tasks that prioritize maximal fidelity and morphological diversity, such as virtual staining or the creation of novel cellular morphotypes, diffusion models currently set the state of the art \cite{TopoCellGen2025CVPR}. Their iterative refinement principle enables precise modeling of complex, multi-scale image distributions, albeit at the cost of substantial computation and slow inference. Conversely, when efficiency is the primary constraint, such as in real-time data augmentation, GANs achieve lower latency through single-pass generation, although this advantage can be offset by training instability and the risk of mode collapse \cite{saad2024gansurvey}. For text generation, autoregressive transformers} \textcolor{black}{excel in producing coherent and contextually consistent pathology narratives. This is primarily because their self-attention mechanism adeptly captures the long-range dependencies required to sustain logical continuity. However, these models remain critically vulnerable to factual hallucination unless rigorously constrained \cite{ji2023hallucinationsurvey}. This landscape of architectural trade-offs underscores that claims of superior performance are contingent not just on the chosen model, but on the specific definitions, datasets, and evaluations employed.}

\subsubsection{Impact on Learning Paradigms}
\textcolor{black}{Generative models are fundamentally reshaping learning paradigms in computational pathology by shifting the objective from learning discriminative boundaries to modeling the underlying data distributions \cite{deshpande2023generative}. This paradigm shift translates into concrete benefits for diagnostic practice, including enhanced data augmentation, improved data efficiency, and richer multimodal representations. First, it redefines \textbf{data augmentation} from simple geometric transformations to semantic synthesis. This is critical in imbalanced tasks, such as mitotic detection or rare cancer subtype classification, where generative models synthesize realistic examples to enhance robustness \cite{xue2021selective}. Second, this focus on the data distribution inherently enables \textbf{data-efficient learning}. Generative objectives support self-supervised pre-training on unlabeled pathology data, yielding versatile backbones. These backbones can be fine-tuned for tasks such as grading, segmentation, and biomarker prediction with limited annotations. Finally, this paradigm enables deeper \textbf{multimodal representations}. By learning generative functions between modalities, models move beyond simple alignment to semantically grounded representations. This provides a more interpretable basis for tasks such as report generation, mutation prediction, and treatment response forecasting \cite{Lu2024}.}

\subsubsection{Methodological Advantages and Clinical Impact}
\textcolor{black}{Generative models derive substantial methodological value by modeling comprehensive data distributions. Their utility arises from three key methodological advantages.
First, by capturing the underlying data distribution, generative models synthesize high-fidelity samples that augment rare classes and reduce annotation demands, thus improving efficiency under data scarcity and imbalance \cite{cai2021generativesurvey}. Second, reconstruction and cross-modal translation objectives yield structured and robust representations that transfer to downstream tasks such as grading, segmentation, and biomarker prediction. They also reveal morphology–molecular associations, providing a basis for hypothesis generation. Third, synthetic cohorts can be shared with reduced risk of exposing protected health information, alleviating privacy barriers and enabling multi-center collaboration.}

\textcolor{black}{Building on these methodological advantages, generative models deliver tangible benefits in clinical and societal impact. In the clinical environment, they enhance operational efficiency. Automated report generation can alleviate the clerical burden on pathologists \cite{Chen_2025_CVPR}, and techniques like virtual staining offer substantial savings in reagent costs and tissue consumption \cite{wang2024cross-modal}. More fundamentally, generative approaches expand core diagnostic capabilities. Virtual molecular profiling enables computational inference of spatially resolved molecular maps from routine H\&E images, unlocking archival resources for biomarker discovery. At the societal level, these models support standardization and broaden access to expertise. Computationally-driven stain normalization can reduce inter-laboratory variability, improving diagnostic consistency for multi-center trials and telepathology. The synthesis of diverse virtual slide libraries, including rare cases, provides a scalable and low-cost resource for global pathology education. By encapsulating specialist knowledge, these models can also function as critical assistive tools in resource-limited settings, thereby helping to mitigate global health disparities.}

\subsection{Critical Barriers to Clinical Translation}
\textcolor{black}{Despite the aforementioned advantages, generative models face enduring barriers to clinical translation. These challenges fall into three categories: intrinsic model limitations, methodological and technical bottlenecks, and systemic obstacles to real-world deployment.
\subsubsection{Intrinsic Model Limitations}
\textbf{Interpretability, Reliability, and Trust Deficit.}
Generative models are trained to approximate data distributions rather than to establish verifiable input–output mappings, which inherently limits their transparency and hampers systematic auditing. Likelihood-driven objectives favor syntactic and visual plausibility over factual accuracy, increasing the risk of hallucinated outputs that compromise interpretation \cite{ji2023hallucinationsurvey}. Domain shifts exacerbate these limitations, introducing instability that undermines robustness across diverse institutional and patient cohorts. In segmentation, models may produce spurious tumor boundaries with misleading precision, potentially misdirecting disease assessment. The high visual fidelity of such errors makes them difficult to detect, unlike obvious artifacts such as blurring, and thereby increases clinical risk. The convergence of limited traceability, poor controllability, and vulnerability to clinical variability jointly creates interpretability challenges that perpetuate institutional mistrust and impede clinical translation. Thus, this opacity raises particular concern in high-stakes diagnostics, where accountability is vital, as it perpetuates a trust deficit that slows clinical adoption.}

\color{black}
\subsubsection{Technical and Evaluation Bottlenecks}
\textbf{WSI Generation Challenges.}
The synthesis of realistic whole-slide images (WSIs) remains a significant challenge due to their gigapixel-scale resolution and the scarcity of comprehensively annotated training data. Current patch-based approaches effectively capture local histological patterns but still struggle to preserve global spatial coherence and tissue-wide architectural integrity. This limitation substantially impairs their utility for applications requiring long-range contextual information, such as tumor staging or morphological heterogeneity assessment. While hierarchical generation frameworks and multiscale attention mechanisms offer promising directions by modeling cross-scale dependencies, they impose prohibitive computational demands that often render them largely impractical on standard hardware \cite{harb2023diffusionbasedgenerationhistopathologicalslide, cechnicka2024urcdm}. Therefore, developing memory-efficient, resolution-aware architectures is a critical research priority. Such advances are essential for modeling complex spatial relationships in histopathological data and moving WSI synthesis toward clinical viability.

\textbf{Evaluation Limitation.}
The absence of standardized, pathology-specific evaluation protocols remains a critical barrier to the clinical adoption of generative models in computational pathology. Conventional metrics such as FID, SSIM, and PSNR, originally developed for natural image assessment, capture only low-level similarity and fail to reflect diagnostic features such as cellular morphology, glandular architecture, or spatial organization \cite{deo2025metricsmatter}. This mismatch risks producing outputs that appear visually plausible but are clinically irrelevant or misleading. Although task-specific evaluation metrics tailored to pathology have been proposed, they remain fragmented and lack systematic validation across diverse datasets and pathological conditions. Furthermore, the field still lacks standardized benchmark datasets and reproducible evaluation pipelines, hindering fair comparison and robust progress \cite{cao2024surveydiffusion}. Addressing these limitations will require the development of pathology-specific benchmarks and multidimensional evaluation frameworks that align with clinical needs.

\subsubsection{Systemic Deployment Obstacles}
\textcolor{black}{\textbf{Barriers to Clinical Deployment.}
The translation of generative models into clinical practice is constrained by systemic barriers that involve technical, regulatory, and socio-technical domains. Technically, state-of-the-art models require substantial computational resources for inference, often exceeding the infrastructure and budgets of typical clinical environments \cite{matheny2022artificial}. Performance degradation due to evolving clinical practices and patient populations necessitates continuous monitoring and periodic retraining, an operational burden that few institutions can sustain. Regulatory and legal uncertainties pose equally significant obstacles. Existing frameworks were built for discriminative algorithms with static performance, whereas generative models produce novel outputs, raising unresolved questions about validation, monitoring, and post-market surveillance \cite{ benjamens2020state}. Liability remains ambiguous, with responsibility for errors caused by generated artifacts not clearly assigned. Finally, socio-technical factors complicate deployment. Human–AI interaction is vulnerable to socio-technical risks, including deployment bias when systems are used beyond their validated scope and automation complacency when clinicians over-rely on algorithmic outputs \cite{drukker2023toward}. These risks are compounded by persistent challenges of clinical validation. Appropriate endpoints for generative systems remain undefined, and large-scale prospective trials required to establish safety and utility are constrained by substantial logistical and financial demands. Confronting these barriers is essential to a credible path toward safe and effective deployment in pathology.}

\textcolor{black}{\textbf{Computational Resource Constraints.}
Generative model development incurs substantial costs across three phases of the computational pipeline. (1) \textbf{Data infrastructure} is burdened by the acquisition, curation, and storage of gigapixel whole-slide images, which demand extensive storage systems and costly expert annotation workflows. (2) \textbf{Training costs} scale exponentially with model complexity \cite{kaplan2020scaling}. Conventional CycleGAN (11.8M parameters) \cite{zhu2017cyclyegan} remains feasible with the computational resources typically available in academic settings, whereas latent diffusion models (1.4B parameters) \cite{rombach2022stablediffusion} and multimodal foundation systems demand multi-GPU training over several days. For instance,  while Quilt-LLaVA \cite{Seyfioglu2024quilt-llava} was trained on LLaVA-v1.5-7B within approximately 10 hours using 4 NVIDIA A100 GPUs, PathChat \cite{Lu2024} required a more resource-intensive multi-stage process, including 32 hours on 32 A100 GPUs for pretraining and an additional 39 hours on 8 A100 GPUs for fine-tuning and joint training.
These comparisons highlight the steep escalation in training costs as models progress from lightweight GANs to diffusion and foundation-scale architectures. (3) \textbf{Deployment costs} persist throughout clinical implementation. These are driven by the substantial computational resources required for high-latency inference, as well as the need for recurrent retraining and monitoring to address data drift and emergent biases.}

\textcolor{black}{These demands create systemic barriers that affect both scientific equity and clinical accessibility. Frontier model development remains largely concentrated within well-resourced institutions, limiting participation from smaller groups and constraining reproducibility \cite{besiroglu2024compute, gundersen2018state}. Clinically, high operational costs restrict deployment in resource-constrained healthcare systems, exacerbating diagnostic disparities and constraining the democratization of advanced pathology tools. Addressing these constraints will require advances in algorithmic efficiency, along with open dissemination of pretrained models and the establishment of standardized benchmarks. Ultimately, the clinical translation of generative pathology depends on reconciling technical innovation with economic sustainability and equitable accessibility.}

\textbf{Ethical, Legal, and Security Considerations.}
The integration of synthetic data in computational pathology raises interconnected challenges across ethical, legal, and security domains \cite{hao2024syntheticchallenges} that must be addressed prior to clinical implementation. Ethically, synthetic histopathology images may appear realistic yet contain biological inconsistencies, risking diagnostic errors by potentially misleading both algorithms and human experts. This concern is particularly acute in high-stakes clinical environments where interpretability and reliability are essential for trust. Legally, the governance of synthetic medical data remains underdeveloped, with unresolved issues spanning patient consent, intellectual property rights, and liability for diagnostic errors. From a security perspective, models trained on synthetic distributions may be less robust to adversarial manipulations, especially when the generated data fail to capture the full heterogeneity and rare edge cases present in clinical specimens.

\subsection{Future Opportunities}
\textcolor{black}{To advance generative models in computational pathology, future efforts must prioritize the deeper integration of data modalities, model capabilities, and clinical needs. This roadmap encompasses unified architectures, multimodal generation, robust evaluation, and enhanced clinical interpretability.}

\textcolor{black}{\textbf{Generative Foundation Model.}
The next stage of generative pathology depend on moving beyond task-specific architectures toward foundation-scale models that can serve as unified engines for diverse clinical and research needs. Current approaches remain fragmented, with models often optimized for narrow objectives or confined to representation learning \cite{zimmermann2024virchow2scalingselfsupervisedmixed}. Generative foundation models (GFMs) are not merely another technical direction but the logical culmination of recent progress, aiming to integrate whole-slide synthesis, mask-conditioned augmentation, semantic reporting, and molecular prediction within a shared and scalable framework. Through multimodal pretraining that aligns histology, genomic profiles, molecular markers, and clinical metadata, GFMs could support cross-modal tasks such as image-to-text, text-to-image, and gene-to-image synthesis. Such integration is crucial for capturing biologically grounded disease representations and for modeling temporal dynamics when sequential histology is combined with longitudinal clinical data. These models could address entrenched challenges of data scarcity, class imbalance, and fragmented pipelines, while enhancing few-shot generalization and cross-domain transferability with minimal supervision. Realizing clinically viable GFMs, however, requires more than technical innovation. It demands strategies to manage the computational burden of foundation-scale training \cite{cao2024surveydiffusion}, ensure semantic consistency across modalities, and establish evaluation and governance frameworks appropriate for clinical contexts.}

\textcolor{black}{\textbf{Agent-driven Generative Pathology}
The emerging field of Agent AI moves beyond static generation toward interactive systems that execute expert workflows through coordinated perception, reasoning, and action \cite{AGENT_AI_SURVEY}. In pathology, this paradigm is already being applied to mirror the diagnostic process, with recent work revealing distinct strategies for action, evidence processing, and report generation \cite{ghezloo2025pathfinder, sun2024pathasst,chen2025evidence}. They exemplify the integration of multi-role coordination, active exploration, and language-based reasoning. The next frontier lies in evolving from passive interpretation to active hypothesis generation and testing. Embedding generative capabilities into the agent loop could transform agents from observers into experimentalists, capable of conducting in silico experiments such as counterfactual simulations of tissue morphology or virtual synthesis of rare diseases. Realizing this vision and ensuring clinical deployability require advances in data, training, and evaluation to capture causal dynamics, unify perception and action, and ensure accurate and robust hypotheses. Maturing these components will enable agent-driven generative pathology to become a dependable partner, capable of delivering insights aligned with the reasoning logic of pathologists.}

\textcolor{black}{\textbf{AI Virtual Cell.}
One long-term direction for generative pathology is the AI Virtual Cell, a multi-scale digital representation that simulates cellular states and perturbation responses in silico \cite{bunne2024virtualcell}. This paradigm shifts the field from static image synthesis to dynamic, hypothesis-driven modeling by leveraging advances in spatial omics and foundational cell atlases \cite{rood2025human, zhang2025systematic}. Virtual cells could serve as priors for generating synthetic tissue at single-cell resolution, with controlled variation in cellular composition and molecular states. They could also enable perturbation-conditioned modeling to simulate disease progression and treatment response, supporting diagnostic training and virtual trials. Furthermore, aligning histology with molecular networks would facilitate mechanistic hypothesis testing, with synthetic microenvironments validated against multimodal ground truths. Realizing this vision requires standardized multi-omics datasets, unified generative–predictive frameworks, and evaluation protocols that assess morphological fidelity, molecular coherence, and causal validity. Together, these advances could transform generative pathology into a hypothesis-aware engine for decoding the cellular basis of disease.}

\textcolor{black}{\textbf{Clinically Deployable Generative Systems.}
The translation of generative models from research prototypes to clinically deployable systems remains a defining challenge. This requires moving beyond isolated performance metrics toward a comprehensive framework that integrates validation and governance. First, a deployable system must demonstrate \textbf{technical robustness}, combining high-fidelity outputs with effective safeguards against factual hallucinations to ensure reliability in clinical use \cite{foote2025embracing}. Such reliability must be sustained under real-world data shifts, a persistent obstacle in practice \cite{zhang2022shifting}. Second, future systems must prove their value through seamless \textbf{clinical integration}, functioning as controllable and interpretable tools for pathologists. Third, these capabilities must be reinforced by a \textbf{governance framework} encompassing regulatory approval pathways, ethical safeguards against bias, and explicit accountability mechanisms \cite{liao2022governance}. Therefore, establishing standards of accuracy, utility, and trustworthiness defines the path toward clinically deployable generative systems in pathology.}

\section{Conclusion}
\textcolor{black}{Content generation modeling} has rapidly gained prominence in computational pathology, becoming a central paradigm for advancing data efficiency, representation learning, and clinical translation. This survey systematically reviewed recent progress by organizing generative approaches into four major task domains: image generation, text generation, \textcolor{black}{molecular profile–morphology generation}, and other specialized generation tasks. The reviewed literature highlights methodological innovations across GANs, diffusion models, and language–vision architectures, showing a clear evolution from proof-of-concept synthesis toward clinically meaningful objectives such as data augmentation, report generation, and biomarker discovery. As the field progresses, content generation foundation models are anticipated to provide scalable, trustworthy, and clinically deployable systems. We hope this work offers both a reference framework and a foundation for future research in this rapidly advancing domain.

\bibliographystyle{IEEEtran} 
\bibliography{reference_new}

\end{document}